\documentclass[AMA,Times1COL]{WileyNJDv5}

\articletype{Research Article}%

\received{Date Sept. 2024}
\revised{Date Month Year}
\accepted{Date Month Year}
\journal{International Journal for Numerical Methods in Engineering}
\volume{126}
\copyyear{2025}

\usepackage{booktabs}
\usepackage{array,multirow,graphicx}
\usepackage{paralist}
\usepackage{xcolor,soul}
\usepackage{caption}
\usepackage{dblfloatfix}
\usepackage{setspace}
\usepackage{svg}
\usepackage{subcaption}
\usepackage{float}

\usepackage{hyperref}
\hypersetup{
    colorlinks=true,
    citecolor=blue,
    linkcolor=blue,
    filecolor=magenta,      
    urlcolor=blue,
    pdfpagemode=FullScreen
    }

\DeclareRobustCommand{\rgamma}{{\mathpalette\irgamma\relax}}
\newcommand{\irgamma}[2]{\raisebox{\depth}{$#1\gamma$}} 

\usepackage{optidef}

\newcommand\finline[3][]{\begin{myfont}[#1]{#2}#3\end{myfont}}%
\newenvironment{myfont}[2][]{\csname#2\endcsname[#1]}{}

\usepackage{silence,lmodern}

\begin{document}

\title{Micropolar elastoplasticity using a fast Fourier transform-based solver}

\author[1,4]{Noah M. Francis}

\author[2]{Ricardo A. Lebensohn}

\author[1,3]{Fatemeh Pourahmadian}

\author[4]{R\'{e}mi Dingreville}

\authormark{FRANCIS \textsc{et al.}}
\titlemark{FRANCIS \textsc{et al.}}

\address[1]{\orgdiv{Department of Civil, Environmental \& Architectural Engineering}, \orgname{University of Colorado Boulder}, \orgaddress{\state{Colorado}, \country{USA}}}

\address[2]{\orgdiv{Theoretical Division}, \orgname{Los Alamos National Laboratory}, \orgaddress{\state{New Mexico}, \country{USA}}}

\address[3]{\orgdiv{Department of Applied Mathematics}, \orgname{University of Colorado Boulder}, \orgaddress{\state{Colorado}, \country{USA}}}

\address[4]{\orgdiv{Center for Integrated Nanotechnologies}, \orgname{Sandia National Laboratories}, \orgaddress{\state{New Mexico}, \country{USA}}}

\corres{Fatemeh Pourahmadian, Department of Civil, Environmental \& Architectural Engineering, University of Colorado Boulder, Colorado, USA. \email{fatemeh.pourahmadian@colorado.edu} \vspace{2mm}\\
R\'{e}mi Dingreville, Center for Integrated Nanotechnologies, Sandia National Laboratories, New Mexico, USA.
\email{rdingre@sandia.gov}}

\fundingInfo{University of Colorado Boulder College of Engineering and Applied Science–Sandia National Laboratories (CEAS-SNL) partnership and Sandia National Laboratories Laboratories Directed Research \& Development program.}

\abstract[Abstract]{
This work presents a micromechanical spectral formulation for obtaining the full-field and homogenized response of elastoplastic micropolar composites. A closed-form radial-return mapping is derived from thermodynamics-based micropolar elastoplastic constitutive equations to determine the increment of plastic strain necessary to return the generalized stress state to the yield surface, and the algorithm implementation is verified using the method of numerically manufactured solutions. Then, size-dependent material response and micro-plasticity are shown as features that may be efficiently simulated in this micropolar elastoplastic framework. The computational efficiency of the formulation enables the generation of large datasets in reasonable computing times.
}


\jnlcitation{\cname{%
\author{Francis NM},
\author{Lebensohn RA},
\author{Pourahmadian F}, and
\author{Dingreville R}}.
\ctitle{Micropolar elastoplasticity using a fast Fourier transform-based solver}. 
\cjournal{\it Int J Numer Methods Eng}. 
\cvol{2025;126:e7651}.
}

\maketitle



\section{Introduction}\label{sec1}
{
Microcontinuum models aim to be physically predictive in situations where the macro-scale and micro-scale are actively coupled. 
Mathematically, this is done by allowing for extra degrees of freedom meant to represent the micro-scale kinematics, independent of the three standard macro-scale displacements.
Maybe the simplest microcontinuum model is the Cosserat/micropolar model introduced in \cite{Cosserat1909,suhubi1964nonlinear,eringen1966linear}, referred to simply as micropolar in this paper.
The micropolar model introduces three additional degrees of freedom called micro-rotations at every point, representing point-wise, rigid-body rotation of the microstructure. {These micro-rotations are caused by moments acting at a point, which are non-existent in the classical theory.}
These additions result in equations that have non-classical effects such as size-dependent mechanical responses that manifest in various ways in solids.
For nanocomposites, size effects appear when surface properties become non-negligible \cite{dingreville2005surface}.
In polycrystals, size effects can emerge as a result of grain size \cite{forest2000cosserat} or due to geometrically necessary dislocations \cite{mayeur2011dislocation}.
Lattice structures also exhibit size effects~\cite{yoder2018size} and the micropolar model has been shown to be a useful effective medium for chiral and non-chiral structures \cite{dos2012construction,spadoni2012elasto, dingreville2014wave, Alberdi2021}.
Numerically, these size effects accounted for by the micropolar model have been shown to regularize softening behavior over different resolutions \cite{Neuner2020}.
}

{
Fast Fourier transform (FFT)-based methods, originally proposed by Moulinec and Suquet \cite{Moulinec1998}, provide an efficient and easy-to-use numerical method for solving partial differential equations governing the response of heterogeneous media such as micropolar media.
In terms of numerical efficiency, the method is advantageous because, at its core, it relies on the FFT algorithm \cite{cooley1965algorithm,frigo2005design}, which has a computational complexity proportional to $n\log(n)$.
This is in contrast to the more standard methods based on Finite Elements, requiring a matrix inversion with complexity typically proportional to $n^{2}$.
The method is easy-to-use because a mesh is not required.
All that is needed as far as the geometry of the heterogeneous medium is concerned is a pixelized (or voxelized) image representing material phase ID in the case of composites, or grain ID in the case of polycrystals.
FFT-based solvers have been used to model, among others problems involving polycrystals and/or composites, the following constitutive behaviors:
(a) elasto-viscoplasticity \cite{Zecevic2022},
(b) dislocation and disclination mechanics \cite{Berbenni2014, djaka2020fft},
(c) strain-gradient plasticity \cite{Lebensohn2016},
(d) couple-stress elasto-viscoplasticity \cite{Upadhyay2016},
(e) elastic wave propagation \cite{Segurado2021}, and
(f) micropolar elasticity \cite{francis2024fast}. The extension of these solvers to non-periodic boundary conditions has also been done \cite{grimm2021fft,risthaus2024fft,risthaus2024imposing}.
This paper extends the authors' work \cite{francis2024fast} to micropolar elastoplasticity using ideas from \cite{russo2020thermomechanics,forest2003elastoviscoplastic,de1993generalisation,steinmann1991localization,simo2006computational}.
}
{
The paper is organized as follows.
In Section \ref{sec:micropolar}, we recall the elastostatic micropolar model \cite{francis2024fast}.
Section \ref{sec:elastoplasticity} presents the micropolar elastoplastic constitutive equations derived in Appendix~\ref{sec:appendixA}, while in Section \ref{sec:radial_return}, we give a closed-form, radial-return mapping algorithm to determine the increment of plastic strain necessary to return the generalized stress state to the yield surface necessary in elastoplastic calculations, derived in Appendix~\ref{sec:closedform}. 
In Section \ref{sec:FFT} we specify the FFT-based solver, which extends \cite{francis2024fast} to elastoplasticity, including the error metrics used in this paper to specify convergence of the method.
In Section \ref{sec:2D}, we exercise the model showing a comparison between the full-field plastic strain maps for both micropolar and classical Cauchy models for different microstructures. In Section \ref{sec:microratcheting}, we show that the model can capture micro-plasticity and dissipation even if the macro-scale remains elastic.
In Section \ref{sec:length_scale}, we illustrate the classical and non-classical limits of this model by varying the constitutive length scales introduced by micropolar mechanics with respect to the geometric length scale.
Finally, in Section \ref{sec:complexity}, we show how our implementation's computing time scales with the input geometry's resolution{, as well as how the nonlinear material parameters affects computing time.} 
}

\subsection{Notation}\label{notation}
{
We use Einstein notation over lowered indices in the standard Cartesian coordinates on $\mathbb{R}^3$ with  
Latin indices $\in \{1,2,3\}$.
A colon before an equals sign ``$:=$'' indicates a definition.
A dot above a symbol indicates a time derivative, and ``$u_{,k}$'' indicates the partial derivative of field $u$ with respect to the $x_k$ spatial coordinate.
}

\section{Methods}\label{sec2}

\subsection{Micropolar mechanics} \label{sec:micropolar}
{
Following Eringen \cite{suhubi1964nonlinear,eringen1966linear,eringen1968theory,Eringen1999}, a micropolar body $\Omega \subset \mathbb{R}^3$ has six degrees of freedom at every point $\mathbf{x}\in \Omega$:
three displacements $u_k(\mathbf{x},t)$ and
three microrotations $\varphi_k(\mathbf{x},t)$. 
Assuming small deformations, we have micropolar strains $e_{kl}(\mathbf{x},t)$ and microcurvature strains $\rgamma_{kl}(\mathbf{x},t)$ defined as, 
\begin{equation}
    \begin{aligned}
    e_{kl} ~:=~ u_{l,k} \,+\, \epsilon_{lkm}\varphi_m \,,\quad\quad\quad \rgamma_{kl} ~:=~ \varphi_{k,l}\,.
    \end{aligned}
    \label{eqn:strains}
\end{equation}
The micropolar balance of linear and angular momentum are then written respectively as
\begin{equation}
    \begin{aligned}
       & t_{kl,k} \,+\, f_l ~=~ 0\,,\\*[0.5mm]
       & m_{kl,k} \,+\, \epsilon_{lkm}t_{km} \,+\, l_l ~=~ 0\,,
        \end{aligned}
        \label{eqn:momemtum_balance}
\end{equation}
where
$t_{kl}(\mathbf{x},t)$ is the stress,
$m_{kl}(\mathbf{x},t)$ is the couple stress,
{$f_l(\mathbf{x},t)$} is the body force per unit volume,
{$l_l(\mathbf{x},t)$} is the body couple per unit volume, and
$\epsilon_{ijk}$ is the Levi-Civita symbol. {Here, $t_{kl}(\mathbf{x},t)$ represents the $l^{\rm th}$ component of the stress vector $\mathbf{t}_k(\mathbf{x},t)$ acting on the positive side of the $k^{\rm th}$ coordinate surface, and similarly for $m_{kl}(\mathbf{x},t)$ (see apter 3 of Eringen \cite{Eringen1962}).}
The possible boundary conditions on an isothermal micropolar body over a time interval $T^+:= [0,\infty)$ can be written as, 
\begin{equation}
     \begin{aligned} 
         & u_k \,=\, \mathfrak{u}_k \quad \text{on} \quad\partial_1\Omega\times T^+\,,\quad\quad \varphi_k \,=\, \phi_k \quad \text{on} \quad\partial_3\Omega\times T^+\,, \\*[1mm]
        & t_{kl}n_k \,=\, \mathcal{T}_l \quad \text{on} \quad\partial_2\Omega\times T^+\,,\quad\quad m_{kl}n_k \,=\, \mathcal{M}_l \quad \text{on} \quad\partial_4\Omega\times T^+\,,
     \end{aligned} 
    \label{eqn:boundary_conditions}
\end{equation}
where $\partial\Omega = \partial_1\Omega\cup\partial_2\Omega = \partial_3\Omega\cup\partial_4\Omega$ and $\partial_1\Omega\cap\partial_2\Omega = \partial_3\Omega\cap\partial_4\Omega = \varnothing$.
{$\mathfrak{u}_k$ and $\phi_k$ correspond to displacement and microrotation boundary conditions, and $\mathcal{T}_l$ and $\mathcal{M}_l$ correspond to force (traction) and moment boundary conditions.}
}

\subsection{Micropolar elastoplasticity}\label{sec:elastoplasticity}
{
This section details the constitutive (elastoplastic) framework used in this paper.
The majority of equations reported here are derived in Appendix~\ref{sec:appendixA}.
Due to linearized kinematics, an elastoplastic additive decomposition of the strains is assumed as
\begin{equation}
    \begin{aligned}
    e_{kl} ~=~ e_{kl}^{\text{e}} \,+\, e_{kl}^{\text{p}}\,,\quad\quad\quad \rgamma_{lk} ~=~ \rgamma_{lk}^{\text{e}} \,+\, \rgamma_{lk}^{\text{p}}\,.
    \end{aligned}
    \label{eqn:additive_decomposition_intext}
\end{equation}
}

{
For centrosymmetric linear elasticity, the constitutive equations are 
\begin{equation}
    \begin{aligned}
    t_{kl} &~=~ A_{klmn}e_{mn}^{\text{e}}\,, \\*[0.5mm]
    m_{kl} &~=~ B_{lkmn}\rgamma_{mn}^{\text{e}}\,,
    \end{aligned}
    \label{eqn:elast_constitutive_aniso}
\end{equation}
with the symmetry conditions $A_{klmn}(\mathbf{x}) = A_{mnkl}(\mathbf{x})$ and $B_{lkmn}(\mathbf{x}) = B_{mnlk}(\mathbf{x})$. {The differing index notations in Eq\@.~\eqref{eqn:elast_constitutive_aniso} stems from the definition of $\rgamma_{kl}(\mathbf{x},t)$ in Eq\@.~\eqref{eqn:strains} as $\varphi_{k,l}(\mathbf{x},t)$ as opposed to $\varphi_{l,k}(\mathbf{x},t)$. The former is common in Eringen's work \cite{Eringen1999}, and the latter is used by Nowacki \cite{nowacki1974linear}.}
When $\Omega$ is isotropic, the fourth-order elastic stiffness tensors can be written as,
\begin{equation}
    \begin{aligned}     
        &   A_{klmn} ~=~ \lambda\delta_{kl}\delta_{mn} \,+\,
                          (\mu + \kappa)\delta_{km}\delta_{ln} \,+\,
                          (\mu - \kappa)\delta_{kn}\delta_{lm}\,, \\*[1mm]
        &   B_{lkmn} ~=~ \alpha\delta_{kl}\delta_{mn} \,+\,
                          \beta\delta_{km}\delta_{ln} \,+\,
                          \gamma\delta_{kn}\delta_{lm}\,,
   \end{aligned}         
   \label{eqn:stiffnesses_iso}
\end{equation}
where $\delta_{kl}$ is the Kronecker delta,
$\lambda(\mathbf{x})$ and $\mu(\mathbf{x})$ are the classical Lam\'{e} parameters ($\lambda$ and $\mu$ measured in Pa), and
$\kappa(\mathbf{x})$, $\alpha(\mathbf{x})$, $\beta(\mathbf{x})$, $\gamma(\mathbf{x})$ are additional micropolar material parameters ($\kappa$ measured in Pa and $\alpha$, $\beta$, $\gamma$ in Pa.m$^2$).
For an isothermal micropolar body $\Omega$ to be thermodynamically stable, we must have the strain-energy-density function positive for all non-zero strains, and equal to zero only when all strains are zero.
This requirement implies the following energetic bounds on the isotropic properties: 
\begin{equation}
    3\lambda \,+\, 2\mu > 0\,, \quad \mu > 0\,, \quad
    \kappa > 0\,, \quad 3\alpha \,+\, \beta \,+\, \gamma > 0\,, \quad
    \gamma \,+\, \beta > 0\,, \quad
    \gamma \,-\, \beta > 0\,.
    \label{eqn:energetic_bounds}
\end{equation}
For completeness, Eq\@.~\eqref{eqn:elast_constitutive_aniso} in its isotropic form is
\begin{equation}
     \begin{aligned} 
        t_{kl} & ~=~ \lambda e_{nn}^{\text{e}}\delta_{kl} \,+\, 2\mu e_{(kl)}^{\text{e}} \,+\, 2\kappa e_{[kl]}^{\text{e}}\,, \\*[1mm]
        m_{kl}   &~=~ \alpha \rgamma_{nn}^{\text{e}}\delta_{kl} \,+\, (\beta + \gamma) \rgamma_{(kl)}^{\text{e}} \,+\, (\beta - \gamma) \rgamma_{[kl]}^{\text{e}}\,,
     \end{aligned} 
    \label{eqn:elast_constitutive_iso}
\end{equation}
where 
\begin{equation}
     \begin{aligned} 
        V_{(kl)} & ~:=~ \frac12\left(V_{kl} + V_{lk}\right)\,, \\*[1mm]
        V_{[kl]} & ~:=~ \frac12\left(V_{kl} - V_{lk}\right)\,,
     \end{aligned} 
    \label{eqn:sym_and_skew}
\end{equation}
are the respective symmetric and skew-symmetric parts of an arbitrary tensor $V_{kl}$.
}

{
For the plasticity model, we first give a multi-criteria yield condition similar to Forest et al. \cite{forest2003elastoviscoplastic} as 
\begin{equation}
    \begin{aligned}
    f(\mathbf{t},p) ~\leq~ 0 \,,\quad\quad g(\mathbf{m},q) ~\leq~ 0\,.
    \end{aligned}
    \label{eqn:yield_conditions_intext}
\end{equation}
where $f\left(\mathbf{t}(\mathbf{x},t), p(\mathbf{x},t)\right)$ and $g\left(\mathbf{m}(\mathbf{x},t), q(\mathbf{x},t)\right)$ are the yield functions of the macro-level and micro-level respectively, and
$p(\mathbf{x},t)$ and $q(\mathbf{x},t)$ are equivalent cumulative plastic strains associated with hardening on the macro-level and micro-level, respectively.
The plastic flow rules implied by Eq\@.~\eqref{eqn:yield_conditions_intext} are given as 
\begin{equation}
    \begin{aligned}
    \dot{e}_{kl}^{\text{p}} ~=~ \lambda_1\frac{\partial f}{\partial t_{kl}} \,,\quad\quad \dot{\rgamma}_{lk}^{\text{p}} ~=~ \lambda_2\frac{\partial g}{\partial m_{kl}}\,,
    \end{aligned}
    \label{eqn:plastic_flow_intext}
\end{equation}
where $\lambda_1(\mathbf{x},t)\,,\lambda_2(\mathbf{x},t) \geq 0$ are two separate plastic multipliers.
}

\subsection{Micropolar radial-return mapping algorithm} \label{sec:radial_return}
{
In this section, we will specify the forms of the yield functions in Eq\@.~\eqref{eqn:yield_conditions_intext} to formulate a micropolar radial-return mapping algorithm.
}

{
Let the body $\Omega$ be isotropic so that we may write the strain energy density as 
\begin{equation}
    \begin{aligned}
     w\left(\mathbf{e}^{\text{e}},\boldsymbol{\rgamma}^{\text{e}},p,q\right) \,:=~ \frac12 e_{kl}^{\text{e}}A_{klmn}e_{mn}^{\text{e}} \,+\, \frac12 \rgamma_{kl}^{\text{e}}B_{klmn}\rgamma_{mn}^{\text{e}} \,+\, \frac12 t_{\text{H}}p^2 \,+\, \frac12 m_{\text{H}}q^2\,,
    \end{aligned}
    \label{eqn:Helmholtz_explicit}
\end{equation}
where
the elastic stiffness tensors $A_{klmn}(\mathbf{x})$ and $B_{klmn}(\mathbf{x})$ are given by Eq\@.~\eqref{eqn:stiffnesses_iso} and
$t_{\text{H}}(\mathbf{x})$ and $m_{\text{H}}(\mathbf{x})$ are material properties that describe isotropic hardening at the macro-level and micro-level, respectively.
The yield function at the macro-level is defined here as
\begin{equation}
    \begin{aligned}
    f(\mathbf{t},p) ~:=~ t_{\text{eq}}(\mathbf{t}) \,-\, R_1(p)\,,
    \end{aligned}
    \label{eqn:yield_function_f}
\end{equation}
where
\begin{equation}
    \begin{aligned}
    t_{\text{eq}}(\mathbf{t}) ~:=\, \sqrt{a_1 s_{(kl)}s_{(kl)} \,+\, a_2 s_{[kl]}s_{[kl]}}\,,
    \end{aligned}
    \label{eqn:equiv_stress}
\end{equation}
is the equivalent stress and $s_{kl}(\mathbf{x},t) := t_{kl}(\mathbf{x},t) - \frac13 t_{nn}(\mathbf{x},t)\delta_{kl}$ are the deviatoric stress components.
The notations $s_{(kl)}$ and $s_{[kl]}$ denote the symmetric and skew-symmetric part of the tensor $s$.
Note that $a_1$ and $a_2$ are dimensionless plastic parameters.
$R_1(p(\mathbf{x},t))$ is the ``radius'' of the macro-level yield surface, defined in the case of linear hardening as 
\begin{equation}
    \begin{aligned}
    R_1(p) ~:=~ t_{\text{Y}} \,+\, t_{\text{H}}p\,,
    \end{aligned}
    \label{eqn:radius_1}
\end{equation}
where $t_{\text{Y}}(\mathbf{x})$ is the macro-level yield stress and the term $t_{\text{H}}p$ describes the hardening. 
The yield function at the micro-level is defined as 
\begin{equation}
    \begin{aligned}
    g(\mathbf{m},q) ~:=~ m_{\text{eq}}(\mathbf{m}) \,-\, R_2(q)\,,
    \end{aligned}
    \label{eqn:yield_function_g}
\end{equation}
where 
\begin{equation}
    \begin{aligned}
    m_{\text{eq}}(\mathbf{m}) ~:=\, \sqrt{b_1 m_{(kl)}m_{(kl)} \,+\, b_2 m_{[kl]}m_{[kl]}}\,,
    \end{aligned}
    \label{eqn:equiv_couple_stress}
\end{equation}
and where 
\begin{equation}
    \begin{aligned}
    R_2(q) ~:=~ m_{\text{Y}} \,+\, m_{\text{H}}q\,,
    \end{aligned}
    \label{eqn:radius_2}
\end{equation}
with $m_{\text{Y}}(\mathbf{x})$ being the micro-level yield stress and the term $m_{\text{H}}q$ describes the hardening at the micro-level.
In the equation above, the units for the plastic parameters $b_1$ and $b_2$ are proportional to the inverse square of a length scale.
There exists many other possibilities for Eqs\@.~\eqref{eqn:equiv_stress},~\eqref{eqn:radius_1},~\eqref{eqn:equiv_couple_stress},~\eqref{eqn:radius_2}.
For various examples, see \cite{russo2020thermomechanics,forest2003elastoviscoplastic,de1993generalisation}.
Using Eqs\@.~\eqref{eqn:yield_function_f}--\eqref{eqn:radius_2} we can now compute
{
\begin{equation}
    \begin{aligned}
    \frac{\partial f}{\partial t_{kl}} ~=~ \frac{a_1s_{(kl)} \,+\, a_2s_{[kl]}}{t_{\text{eq}}(\mathbf{t})}\,,
    \end{aligned}
    \label{eqn:dfdt}
\end{equation}
}
and 
{
\begin{equation}
    \begin{aligned}
    \frac{\partial g}{\partial m_{kl}} ~=~ \frac{b_1m_{(kl)} \,+\, b_2m_{[kl]}}{m_{\text{eq}}(\mathbf{m})}\,,
    \end{aligned}
    \label{eqn:dgdm}
\end{equation}
}
which are used in the flow rules defined in Eq\@.~\eqref{eqn:plastic_flow_intext}.
Similar to de Borst \cite{de1993generalisation}, the equivalent plastic strain rates $\dot{p}(\mathbf{x},t)$ and $\dot{q}(\mathbf{x},t)$ may be defined using the plastic part of the strain rate as 
{
\begin{equation}
    \begin{aligned}
    \dot{p} ~:=~ \sqrt{\frac{1}{a_1}\dot{e}_{(kl)}^{\text{p}}\dot{e}_{(kl)}^{\text{p}} \,+\, \frac{1}{a_2}\dot{e}_{[kl]}^{\text{p}}\dot{e}_{[kl]}^{\text{p}}}\,,
    \end{aligned}
    \label{eqn:p_dot}
\end{equation}
}
and
{
\begin{equation}
    \begin{aligned}
    \dot{q} ~:=~ \sqrt{\frac{1}{b_1}\dot{\rgamma}_{(lk)}^{\text{p}}\dot{\rgamma}_{(lk)}^{\text{p}} \,+\, \frac{1}{b_2}\dot{\rgamma}_{[lk]}^{\text{p}}\dot{\rgamma}_{[lk]}^{\text{p}}}\,,
    \end{aligned}
    \label{eqn:q_dot}
\end{equation}
}
such that substituting Eq\@.~\eqref{eqn:plastic_flow_intext} into Eqs\@.~\eqref{eqn:p_dot} and~\eqref{eqn:q_dot} results in
\begin{equation}
    \begin{aligned}
    \dot{p} ~=~ \lambda_1\,,\quad\quad \dot{q} ~=~ \lambda_2\,.
    \end{aligned}
    \label{eqn:pq_dot_equals_lam12}
\end{equation}

With these definitions, we can solve for elastoplastic stresses in the strain-driven loading setting. We begin with the stress-rate equations found by taking the time derivative of Eq\@.~\eqref{eqn:elast_constitutive_aniso} and applying Eq\@.~\eqref{eqn:additive_decomposition_intext}
\begin{equation}
    \begin{aligned}
    \dot{t}_{kl} &~=~ A_{klmn}\dot{e}_{mn}^{\text{e}} ~=~ A_{klmn}\left(\dot{e}_{mn} - \dot{e}_{mn}^{\text{p}}\right)\,,\\*[1mm]
    \dot{m}_{kl} &~=~ B_{lkmn}\dot{\rgamma}_{mn}^{\text{e}} ~=~ B_{lkmn}\left(\dot{\rgamma}_{mn} - \dot{\rgamma}_{mn}^{\text{p}}\right)\,.
    \end{aligned}
    \label{eqn:stress_rates}
\end{equation}
Replacing Eqs\@.~\eqref{eqn:plastic_flow_intext} and~\eqref{eqn:dfdt}--\eqref{eqn:q_dot} in the plastic flow rules, we get 
{
\begin{equation}
    \begin{aligned}
    \dot{t}_{kl} &~=~ A_{klmn}\dot{e}_{mn} \,-\, \frac{\dot{p}}{t_{\text{eq}}(\mathbf{t})}\left(a_1A_{klmn}s_{(mn)} \,+\, a_2A_{klmn}s_{[mn]}\right)\,,\\*[1mm]
    \dot{m}_{kl} &~=~ B_{lkmn}\dot{\rgamma}_{mn} \,-\, \frac{\dot{q}}{m_{\text{eq}}(\mathbf{m})}\left(b_1B_{lkmn}m_{(nm)} \,+\, b_2B_{lkmn}m_{[nm]}\right)\,.
    \end{aligned}
    \label{eqn:stress_rates_flow_rules_plugged_in}
\end{equation}
}
These equations will be discretized using the Euler-Backward (EB) implicit numerical integration scheme.
Applying EB to some first-order ordinary differential equation, $\dot{x}(t) = F\left(x(t),t\right)$, gives the following implicit algebraic equation for $x(t_{n+1})$
\begin{equation}
    \begin{aligned}
    \frac{x\left(t_{n+1}\right) - x\left(t_{n}\right)}{t_{n+1} - t_{n}} ~=~ F\left(x\left(t_{n+1}\right), t_{n+1}\right)\,.
    \end{aligned}
    \label{eqn:EB_1}
\end{equation}
To simplify this notation, let us write
\begin{equation}
    \begin{aligned}
    \frac{x^{(n+1)} - x^{(n)}}{t_{n+1} - t_{n}} ~=~ F\left(x^{(n+1)}, t_{n+1}\right)\,,
    \end{aligned}
    \label{eqn:EB_2}
\end{equation}
and simplifying even more, we can drop the superscripted reference to $(n+1)$ and write
\begin{equation}
    \begin{aligned}
    \frac{x - x^{(n)}}{t_{n+1} - t_{n}} ~=~ F(x, t_{n+1})\,.
    \end{aligned}
    \label{eqn:EB_3}
\end{equation}
Applying EB to the stress rate equations in  Eq\@.~\eqref{eqn:stress_rates_flow_rules_plugged_in} using the notation in Eq\@.~\eqref{eqn:EB_3}, we get
{
\begin{equation}
    \begin{aligned}
    t_{kl} - t_{kl}^{(n)}  &~=~ A_{klmn}\left(e_{mn} - e_{mn}^{(n)}\right) \,-\, \frac{p - p^{(n)}}{t_{\text{eq}}\left(\mathbf{t}\right)}\left(a_1A_{klmn}s_{(mn)} \,+\, a_2A_{klmn}s_{[mn]}\right)\,,\\*[1mm]
    m_{kl} - m_{kl}^{(n)} &~=~ B_{lkmn}\left(\rgamma_{mn} - \rgamma_{mn}^{(n)}\right) \,-\, \frac{q - q^{(n)}}{m_{\text{eq}}(\mathbf{m})}\left(b_1B_{lkmn}m_{(nm)} \,+\, b_2B_{lkmn}m_{[nm]}\right)\,,
    \end{aligned}
    \label{eqn:stress_rates_flow_rules_plugged_in_EB_3}
\end{equation}
}
noting that $A_{klmn}(\mathbf{x})$, $B_{lkmn}(\mathbf{x})$, $a_1(\mathbf{x})$, $a_2(\mathbf{x})$, $b_1(\mathbf{x})$ and $b_2(\mathbf{x})$ are all time-invariant.
For strain-driven loading, let us assume we know all fields at time $t_n$, as well as the strains $e_{kl}(\mathbf{x})$ and $\rgamma_{lk}(\mathbf{x})$ at time $t_{n+1}$, and thus need to solve for the stresses $t_{kl}(\mathbf{x})$ and $m_{kl}(\mathbf{x})$ at time $t_{n+1}$.
Define the elastic trial stresses at time $t_{n+1}$ as
\begin{equation}
    \begin{aligned}
    t_{kl}^{\text{trial}} &\,:=~ t_{kl}^{(n)} \,+\, A_{klmn}\left(e_{mn} - e_{mn}^{(n)}\right)\,,\\*[1mm]
    m_{kl}^{\text{trial}} &\,:=~ m_{kl}^{(n)} \,+\, B_{lkmn}\left(\rgamma_{mn} - \rgamma_{mn}^{(n)}\right)\,.
    \end{aligned}
    \label{eqn:trial_stresses}
\end{equation}
These are known quantities, and allow us to write Eq\@.~\eqref{eqn:stress_rates_flow_rules_plugged_in_EB_3} as
{
\begin{equation}
    \begin{aligned}
    t_{kl}  &~=~ t_{kl}^{\text{trial}} \,-\, \frac{p - p^{(n)}}{t_{\text{eq}}\left(\mathbf{t}\right)}\left(a_1A_{klmn}s_{(mn)} \,+\, a_2A_{klmn}s_{[mn]}\right)\,,\\*[1mm]
    m_{kl} &~=~ m_{kl}^{\text{trial}} \,-\, \frac{q - q^{(n)}}{m_{\text{eq}}(\mathbf{m})}\left(b_1B_{lkmn}m_{(nm)} \,+\, b_2B_{lkmn}m_{[nm]}\right)\,.
    \end{aligned}
    \label{eqn:stress_rates_flow_rules_plugged_in_EB_3_trial}
\end{equation}
}
}

{
The unknowns at time $t_{n+1}$ are the stresses $t_{kl}(\mathbf{x})$, $m_{kl}(\mathbf{x})$ and the plastic strains $p(\mathbf{x})$, $q(\mathbf{x})$. In the current framework, we are able to solve for the plastic strains at time $t_{n+1}$ as closed-form functions of the material parameters and the known elastic trial stresses $t_{kl}^{\text{trial}}(\mathbf{x})$, $m_{kl}^{\text{trial}}(\mathbf{x})$ at time $t_{n+1}$. This closed-form update, derived in Appendix~\ref{sec:closedform}, is given here as
{
\begin{equation}
    \begin{aligned}
    p &~=~ p^{(n)}\frac{2a_1\mu}{t_{\text{H}} + 2a_1\mu} \,+\, \frac{t_{\text{eq}}(\mathbf{t}^{\text{trial}}) - t_{\text{Y}}}{t_{\text{H}} + 2a_1\mu}\,,\\*[1mm]
    q &~=~ q^{(n)}\frac{b_1(\gamma + \beta)}{m_{\text{H}} + b_1(\gamma + \beta)} \,+\, \frac{m_{\text{eq}}(\mathbf{m}^{\text{trial}}) - m_{\text{Y}}}{m_{\text{H}} + b_1(\gamma+\beta)}\,.
    \end{aligned}
    \label{eqn:trial_stresses2}
\end{equation}
}
Owing to Eq\@.~\eqref{eqn:trial_stresses2}, we may use Eq\@.~\eqref{eqn:stress_rates_flow_rules_plugged_in_EB_3_trial} as the update for the the stresses $t_{kl}(\mathbf{x})$, $m_{kl}(\mathbf{x})$. The algorithm for the micropolar radial-return mapping is now given in Alg\@.~\ref{alg:radial_return}. 

\begin{algorithm}
\caption{\enskip Micropolar Radial-Return Mapping}\label{alg:radial_return}
\setstretch{1.55}
\begin{algorithmic}
    \State \textbf{Input:} $e_{kl}$, $\rgamma_{lk}$, $e_{kl}^{(n)}$, $\rgamma_{lk}^{(n)}$, $t_{kl}^{(n)}$, $m_{kl}^{(n)}$, $p^{(n)}$, $q^{(n)}$, and all constitutive parameters.
    \State \textbf{Output:} $t_{kl}$, $m_{kl}$, $p$, $q$.
    \State Compute $t_{kl}^{\text{\finline[]{Times}{trial}}}$ and $m_{kl}^{\text{\finline[]{Times}{trial}}}$ using Eq\@.~\eqref{eqn:trial_stresses}.
    \State Set $a_2 = a_1\frac{\mu}{\kappa}$ and $b_2 = b_1\frac{\gamma+\beta}{\gamma-\beta}$.
    \State Compute $t_{\text{\finline[]{Times}{eq}}}(\mathbf{t}^{\text{\finline[]{Times}{trial}}})$ using Eq\@.~\eqref{eqn:equiv_stress} and $m_{\text{\finline[]{Times}{eq}}}(\mathbf{m}^{\text{\finline[]{Times}{trial}}})$ using Eq\@.~\eqref{eqn:equiv_couple_stress}.
    \If{$t_{\text{\finline[]{Times}{eq}}}(\mathbf{t}^{\text{\finline[]{Times}{trial}}}) < t_{\text{\finline[]{Times}{Y}}} + t_{\text{\finline[]{Times}{H}}}p^{(n)}$}
    \State Set $p = p^{(n)}$ and $t_{kl} = t_{kl}^{\text{\finline[]{Times}{trial}}}$. \Comment{macro-elastic}
    \Else
    \State Compute $p$ using Eq\@.~\eqref{eqn:trial_stresses2}.
    \State Set {$s_{(kl)} = \left(\frac{t_{\text{\finline[]{Times}{Y}}} \,+\, t_{\text{\finline[]{Times}{H}}}p}{t_{\text{\finline[]{Times}{Y}}} \,+\, t_{\text{\finline[]{Times}{H}}}p \,+\, 2(p - p^{(n)})a_1\mu}\right)s_{(kl)}^{\text{\finline[]{Times}{trial}}}$} and {$s_{[kl]} = \left(\frac{t_{\text{\finline[]{Times}{Y}}} \,+\, t_{\text{\finline[]{Times}{H}}}p}{t_{\text{\finline[]{Times}{Y}}} \,+\, t_{\text{\finline[]{Times}{H}}}p \,+\, 2(p - p^{(n)})a_2\kappa}\right)s_{[kl]}^{\text{\finline[]{Times}{trial}}}$}.
    \State Set $t_{kl} = \frac13 t_{nn}^{\text{\finline[]{Times}{trial}}}\delta_{kl} \,+\,  s_{(kl)} \,+\, s_{[kl]}$. \Comment{macro-plastic}
    \EndIf
    \If{$m_{\text{\finline[]{Times}{eq}}}(\mathbf{m}^{\text{\finline[]{Times}{trial}}}) < m_{\text{\finline[]{Times}{Y}}} + m_{\text{\finline[]{Times}{H}}}q^{(n)}$}
    \State Set $q = q^{(n)}$ and $m_{kl} = m_{kl}^{\text{\finline[]{Times}{trial}}}$. \Comment{micro-elastic}
    \Else
    \State Compute $q$ using Eq\@.~\eqref{eqn:trial_stresses2}.
    \State Set {$m_{(kl)} = \left(\frac{m_{\text{\finline[]{Times}{Y}}} \,+\, m_{\text{\finline[]{Times}{H}}}q}{m_{\text{\finline[]{Times}{Y}}} \,+\, m_{\text{\finline[]{Times}{H}}}q \,+\, \left(q - q^{(n)}\right)b_1(\gamma + \beta)}\right)m_{(kl)}^{\text{\finline[]{Times}{trial}}}$} and {$m_{[kl]} = \left(\frac{m_{\text{\finline[]{Times}{Y}}} \,+\, m_{\text{\finline[]{Times}{H}}}q}{m_{\text{\finline[]{Times}{Y}}} \,+\, m_{\text{\finline[]{Times}{H}}}q \,+\, \left(q - q^{(n)}\right)b_2(\gamma - \beta)}\right)m_{[kl]}^{\text{\finline[]{Times}{trial}}}$}.
    \State Set $m_{kl} = m_{(kl)} \,+\, m_{[kl]}$. \Comment{micro-plastic}
    \EndIf
\end{algorithmic}
\end{algorithm}
}

\subsection{Micropolar elastoplastic FFT-based solver}\label{sec:FFT}
{
{We used a} FFT-based algorithm {which} is a version of the Moulinec-Suquet ``basic scheme'' with time-stepping, to account for the history dependence that the elastoplastic framework requires.
Using both Algorithm 1 from \cite{francis2024fast} and Alg\@.~\ref{alg:radial_return} in the present work, the logic for the micropolar elastoplastic FFT-based solver, assuming $Q_{klmn}(\mathbf{x}) = 0$ for all $\mathbf{x}\in\Omega$, is given in Alg\@.~\ref{alg:fft}. {This fixed-point algorithm numerically solves the total strain and rotations integral equations (called the \textit{micropolar Lippmann-Schwinger equations})
\begin{equation}
    \begin{aligned}
    e_{kl}(\mathbf{x},t) ~&=~ E_{kl}(t) \,+\, \int_{\Omega}\left(\Gamma_{klmn}^1(\mathbf{x}-\mathbf{x}')\tau_{mn}(\mathbf{x}',t) \,+\, \Gamma_{klmn}^2(\mathbf{x}-\mathbf{x}')\mu_{mn}(\mathbf{x}',t) \,+\, \Lambda_{klmn}^1(\mathbf{x}-\mathbf{x}')\tau_{mn}^0(\mathbf{x}',t)\right)\text{d}\mathbf{x}'\,, \\*[1mm]
    \rgamma_{kl}(\mathbf{x},t) ~&=~ \Gamma_{kl}(t) \,+\, \int_{\Omega}\left(\Gamma_{klmn}^3(\mathbf{x}-\mathbf{x}')\tau_{mn}(\mathbf{x}',t) \,+\, \Gamma_{klmn}^4(\mathbf{x}-\mathbf{x}')\mu_{mn}(\mathbf{x}',t) \,+\, \Lambda_{klmn}^2(\mathbf{x}-\mathbf{x}')\tau_{mn}^0(\mathbf{x}',t)\right)\text{d}\mathbf{x}'\,,
    \end{aligned}
\label{eqn:micropolar_LippmannSchwinger}
\end{equation}
for the total strains $e_{kl}(\mathbf{x},t)$ and $\rgamma_{kl}(\mathbf{x},t)$ under prescribed average strains $E_{kl}(t)$ and $\Gamma_{kl}(t)$.
The tensors $\Gamma^{i}_{klmn}(\mathbf{x})$ for $i\in\{1,\dots,4\}$ and $\Lambda^{j}_{klmn}(\mathbf{x})$ for $j\in\{1,2\}$ are Green’s tensors required for the solution of the micropolar problem, defined in \cite{francis2024fast}.
It is worth noting that in the linear elastic case\cite{francis2024fast}, Eq\@.~\eqref{eqn:micropolar_LippmannSchwinger} is a set of linear integral equations due to the polarization stresses being linearly related to the total strains. 
However, for elastoplasticity (and other nonlinear constitutive choices) these equations are now \textit{nonlinear} integral equations due to the polarization stresses being nonlinearly related to the total strains. In both cases, the polarization stresses are the difference between the actual generalized
stress and the generalized stress in a linear reference medium, defined as 
\begin{equation}
    \begin{aligned}
    \tau_{kl}(\mathbf{x},t) ~&:=~ t_{kl}(\mathbf{x},t) \,-\, A^0_{klmn}e_{mn}(\mathbf{x},t)\,, \\*[1mm]
    \mu_{kl}(\mathbf{x},t) ~&:=~ m_{kl}(\mathbf{x},t) \,-\, B^0_{lkmn}\rgamma_{mn}(\mathbf{x},t)\,, \\*[1mm]
    \tau_{kl}^0(\mathbf{x},t) ~&:=\, \tau_{kl}(\mathbf{x},t) \,+\, A^0_{klmn}E_{mn}(t)\,,
    \end{aligned}
\label{eqn:micropolar_polarizatons}
\end{equation}
where the tensors $A^0_{klmn}$ and $B^0_{lkmn}$ are micropolar elasticity tensors of a homogeneous reference medium and $\tau_{kl}^0$ is a polarization-like stress tensor.
%
%
For numerically computing the fixed-point iterations of Eq\@.~\eqref{eqn:micropolar_LippmannSchwinger}, the convolution theorem along with FFTs are used, resulting in the following iteration equation
\begin{equation}
    \begin{aligned}
    e_{kl}^{(i+1)}(\mathbf{x},t) ~&=~ E_{kl}(t) \,+\, \mathcal{F}^{-1}\left[\hat{\Gamma}_{klmn}^1(\mathbf{k})\hat{\tau}_{mn}^{(i)}(\mathbf{k},t) \,+\, \hat{\Gamma}_{klmn}^2(\mathbf{k})\hat{\mu}_{mn}^{(i)}(\mathbf{k},t) \,+\, \hat{\Lambda}_{klmn}^1(\mathbf{k})\hat{\tau}_{mn}^{0(i)}(\mathbf{k},t)\right](\mathbf{x},t)\,, \\*[1mm]
    \rgamma_{kl}^{(i+1)}(\mathbf{x},t) ~&=~ \Gamma_{kl}(t) \,+\, \mathcal{F}^{-1}\left[\hat{\Gamma}_{klmn}^3(\mathbf{k})\hat{\tau}_{mn}^{(i)}(\mathbf{k},t) \,+\, \hat{\Gamma}_{klmn}^4(\mathbf{k})\hat{\mu}_{mn}^{(i)}(\mathbf{k},t) \,+\, \hat{\Lambda}_{klmn}^2(\mathbf{k})\hat{\tau}_{mn}^{0(i)}(\mathbf{k},t)\right](\mathbf{x},t)\,,
    \end{aligned}
\label{eqn:micropolar_LippmannSchwinger_iteration}
\end{equation}
where $e_{kl}^{(0)}(\mathbf{x},t) = E_{kl}(t)$ and $\rgamma_{kl}^{(0)}(\mathbf{x},t) = \Gamma_{kl}(t)$, and $\mathcal{F}^{-1}$ corresponds to the inverse Fourier Transform.
}

{
Here, the notation $\hat{\cdot}$ indicates that a variable is Fourier-transformed, so $\hat{\Gamma}^1_{klmn}(\mathbf{k})$, $\hat{\Gamma}^2_{klmn}(\mathbf{k})$, $\hat{\Gamma}^3_{klmn}(\mathbf{k})$, $\hat{\Gamma}^4_{klmn}(\mathbf{k})$, $\hat{\Lambda}^1_{klmn}(\mathbf{k})$, $\hat{\Lambda}^2_{klmn}(\mathbf{k})$ are the Fourier-transformed Green's tensors given in \cite{francis2024fast}.
These may either be pre-computed and stored, or computed on the fly when needed.
While {many} convergence criteria exist\cite{ferrier2023posteriori}, the error measure $err^{(i+1)}$ is computed here as 
\begin{equation}
    err^{(i+1)} ~=~ \max\left(err_{\mathbf{e}}^{(i+1)}, err_{\mathbf{t}}^{(i+1)}, err_{\boldsymbol{\rgamma}}^{(i+1)}, err_{\mathbf{m}}^{(i+1)}\right) \,,
\label{eqn:err}
\end{equation}
where we have either 
\begin{equation}
    err^{(i+1)}_{\boldsymbol{a}} ~:=~ \frac{\left\langle\left\|\boldsymbol{a}^{(n,i+1)}(\mathbf{x}) - \boldsymbol{a}^{(n,i)}(\mathbf{x})\right\|\right\rangle_{\Omega}}{\left\|\left\langle\boldsymbol{a}^{(n,i+1)}(\mathbf{x})\right\rangle_{\Omega}\right\|}\,,
\label{eqn:err_average}
\end{equation}
or 
\begin{equation}
    err^{(i+1)}_{\boldsymbol{a}} ~:=~ \frac{\max\limits_{\mathbf{x}}\left\|\boldsymbol{a}^{(n,i+1)}(\mathbf{x}) - \boldsymbol{a}^{(n,i)}(\mathbf{x})\right\|}{\left\|\left\langle\boldsymbol{a}^{(n,i+1)}(\mathbf{x})\right\rangle_{\Omega}\right\|}\,,
\label{eqn:err_local}
\end{equation}
with $\boldsymbol{a} \in \{\mathbf{e},\mathbf{t},\boldsymbol{\rgamma},\mathbf{m}\}$ and $\|\mathbf{V}\| := \sqrt{V_{kl}V_{kl}}$.
Using Eq\@.~\eqref{eqn:err_average} will result in a solution that has converged under the threshold value $\varepsilon$ on average, while using Eq\@.~\eqref{eqn:err_local} ideally results in a ``pixel-wise'' or locally converged solution.
}
\begin{algorithm}
\caption{\enskip Micropolar Elastoplastic FFT-based Basic Scheme}\label{alg:fft}
\setstretch{1.55}
\begin{algorithmic}
    \State \textbf{Input:} $A_{klmn}(\mathbf{x})$, $B_{lkmn}(\mathbf{x})$, $t_\text{\finline[]{Times}{Y}}(\mathbf{x})$, $t_\text{\finline[]{Times}{H}}(\mathbf{x})$, $m_\text{\finline[]{Times}{Y}}(\mathbf{x})$, $m_\text{\finline[]{Times}{H}}(\mathbf{x})$, $E_{kl}(t)$, $\Gamma_{kl}(t)$, $\varepsilon$ \\  $\forall t\in T^+:=(t_1,t_2,...,t_n,t_{n+1},...,t_N,t_{N+1})$ and $\forall \mathbf{x}\in\Omega$.
    \State \textbf{Output:} $e_{kl}(\mathbf{x},t)$, $\rgamma_{kl}(\mathbf{x},t)$, $t_{kl}(\mathbf{x},t)$, $m_{kl}(\mathbf{x},t)$, $p(\mathbf{x},t)$, $q(\mathbf{x},t)$ $\forall t\in T^+$ and $\forall \mathbf{x}\in\Omega$.
    \State Compute $A_{klmn}^0 = \langle A_{klmn}(\mathbf{x})\rangle_{\Omega}$ and $B_{lkmn}^0 = \langle B_{lkmn}(\mathbf{x})\rangle_{\Omega}$.
    \State Set $e_{kl}^{(1)}(\mathbf{x}) = \rgamma_{kl}^{(1)}(\mathbf{x}) = t_{kl}^{(1)}(\mathbf{x}) = m_{kl}^{(1)}(\mathbf{x}) = 0$, $p^{(1)}(\mathbf{x}) = q^{(1)}(\mathbf{x}) = 0$ $\forall\mathbf{x}\in\Omega$.
    \For{$n\in(2,\dots,N+1)$} 
    \If{$n=2$}
    \State Set $e_{kl}^{(n,1)} = E_{kl}^{(n)}$ and $\rgamma_{kl}^{(n,1)} = \Gamma_{kl}^{(n)}$ $\forall\mathbf{x}\in\Omega$.
    \State Compute $t_{kl}^{(n,1)}$, $m_{kl}^{(n,1)}$, $p^{(n,1)}$, and $q^{(n,1)}$ using Alg~\ref{alg:radial_return}.
    \Else
    \State Set $e_{kl}^{(n,1)} = e_{kl}^{(n-1)} \,+\, \frac{t_n \,-\, t_{n-1}}{t_{n-1} \,-\, t_{n-2}}\left(e_{kl}^{(n-1)} - e_{kl}^{(n-2)}\right)$ \\ \ \ \ \ \ \ \ and $\rgamma_{kl}^{(n,1)} = \rgamma_{kl}^{(n-1)} \,+\, \frac{t_n \,-\, t_{n-1}}{t_{n-1} \,-\, t_{n-2}}\left(\rgamma_{kl}^{(n-1)} - \rgamma_{kl}^{(n-2)}\right)$ $\forall\mathbf{x}\in\Omega$. \Comment{linear extrapolation}
    \State Compute $t_{kl}^{(n,1)}$, $m_{kl}^{(n,1)}$, $p^{(n,1)}$, and $q^{(n,1)}$ using Alg~\ref{alg:radial_return}.
    \EndIf
    \State Set $err^{(1)} = 2\varepsilon$ and $T_{kl}^0 = A_{klmn}^0E_{mn}^{(n)}$.
    \While{$err^{(i)} > \varepsilon$}
    \State Set polarization stresses: $\tau_{kl}^{(i)} = t_{kl}^{(n,i)} \,-\, A_{klmn}^0 e_{mn}^{(n,i)}$, $\mu_{kl}^{(i)} = m_{kl}^{(n,i)} \,-\, B_{lkmn}^0 \rgamma_{mn}^{(n,i)}$, and $\tau_{kl}^{0(i)} = \tau_{kl}^{(i)} \,+\, T_{kl}^0$.
    \State Fourier transform: $\hat{\tau}_{kl}^{(i)}=\mathcal{F}\left[\tau_{kl}^{(i)}\right]$, $\hat{\mu}_{kl}^{(i)}=\mathcal{F}\left[\mu_{kl}^{(i)}\right]$, and  $\hat{\tau}_{kl}^{0(i)}=\mathcal{F}\left[\tau_{kl}^{0(i)}\right]$.
    \State Contract with Green's tensors: \\ \ \ \ \ \ \ \ \ \ \ $\hat{\tilde{e}}_{kl}^{(n,i+1)} = \hat{\Gamma}^1_{klmn}\hat{\tau}_{mn}^{(i)} \,+\, \hat{\Gamma}^2_{klmn}\hat{\mu}_{mn}^{(i)} \,+\, \hat{\Lambda}^1_{klmn}\hat{\tau}_{mn}^{0(i)}$ and $\hat{\tilde{\rgamma}}_{kl}^{(n,i+1)} = \hat{\Gamma}^3_{klmn}\hat{\tau}_{mn}^{(i)} \,+\, \hat{\Gamma}^4_{klmn}\hat{\mu}_{mn}^{(i)} \,+\, \hat{\Lambda}^2_{klmn}\hat{\tau}_{mn}^{0(i)}$.
    \State Inverse Fourier transform: $e_{kl}^{(n,i+1)}=\mathcal{F}^{-1}\left[\hat{\tilde{e}}_{kl}^{(n,i+1)}\right] + E_{kl}^{(n)}$ and $\rgamma_{kl}^{(n,i+1)}=\mathcal{F}^{-1}\left[\hat{\tilde{\rgamma}}_{kl}^{(n,i+1)}\right] + \Gamma_{kl}^{(n)}$.
    \State Compute $t_{kl}^{(n,i+1)}$, $m_{kl}^{(n,i+1)}$, $p^{(n,i+1)}$, and $q^{(n,i+1)}$ using Alg~\ref{alg:radial_return}.
    \State Compute $err^{(i+1)}$ using Eq\@.~\eqref{eqn:err}.
    \EndWhile
    \State Set $\left(e_{kl}^{(n)}, \rgamma_{kl}^{(n)}, t_{kl}^{(n)}, m_{kl}^{(n)}, p^{(n)}, q^{(n)}\right) = \left(e_{kl}^{(n,\text{\finline[]{Times}{last})}}, \rgamma_{kl}^{(n,\text{\finline[]{Times}{last})}}, t_{kl}^{(n,\text{\finline[]{Times}{last})}}, m_{kl}^{(n,\text{\finline[]{Times}{last})}}, p^{(n,\text{\finline[]{Times}{last})}}, q^{(n,\text{\finline[]{Times}{last})}}\right)$.
    \EndFor
\end{algorithmic}
\end{algorithm}


\section{Results}

\subsection{Plastic strain examples in two dimensions}
\label{sec:2D}
{
The results presented in Appendix~\ref{sec:verification} establish credence in the correctness and accuracy of our FFT-based micropolar elastoplasticity implementation.
In this subsection, we exercise this algorithm to illustrate how the results of our micropolar elastoplastic model contrasts with a classical Cauchy continuum formulation. 
}

{
To demonstrate Alg\@.~\ref{alg:fft} in action, we plot in Fig\@.~\ref{fig:plastic_strain} the plastic strain $p(x_1,x_2,t=0.64)$ and the averaged stress-strain curve for both the classical Cauchy and micropolar continua.
{Response of the matrix phase (softer than inclusion) is also included for comparison.}
We exercised both models over four example geometries:
an off-centered circular inclusion,
five circular inclusions of varying radii randomly distributed within a matrix,
ten circular inclusions of equal radius randomly distributed within a matrix, and
a spinodal decomposition microstructure.
The geometries with the circular inclusions were generated manually, while the spinodal decomposition microstructure was generated via the phase-field method~\cite{Stewart2020, dingreville2020benchmark}.
Starting from the natural zero-stress/zero-strain state, we loaded those geometries monotonically for $N = 100$-time steps using a $\Delta t = 0.01$.
For the micropolar strain rate, we used $\dot{E}_{12}(t) = 1$, and for the Cauchy strain rate we used $\dot{E}_{12}(t) = \dot{E}_{21}(t) = 1$, with the other components set to zero in both cases.
In the geometries shown in Fig\@.~\ref{fig:plastic_strain}, black represents phase 1 and white represents phase 2.
The micropolar constitutive parameters are listed in Table~\ref{table:micropolar-res1}. {The error threshold of the FFT-based iterative algorithm is $\varepsilon = 1\times10^{-6}$ with the error metric defined in Eq\@.~\eqref{eqn:err_local}}.
}

\begin{table*}[!ht]%
\centering
\caption{Micropolar constitutive parameters of phases 1 and 2 plus parameters chosen for classical Cauchy limit.}
\label{table:micropolar-res1}
\begin{tabular*}{\textwidth}{@{\extracolsep\fill}lllllllllllll@{\extracolsep\fill}}%
\toprule
\textbf{Parameter} & $\lambda$ & $\mu$ & $\kappa$ & $\alpha$ & $\beta$ & $\gamma$ & $t_{\text{Y}}$ & $t_{\text{H}}$ & $m_{\text{Y}}$ & $m_{\text{H}}$ & $a_1$ & $b_1$ \\
\midrule
\textbf{Phase 1}              & 1.0 & 1.0 & 1.0    & 0.0 & 0.0 & 1.0    & 0.5  & {0.125} & 0.5    & {0.125} & 1.5 & 1.5   \\
\textbf{Phase 2}              & 2.0 & 2.0 & 2.0    & 0.0 & 0.0 & 2.0    & 0.75 & {0.25}  & 0.75   & {0.25}  & 1.5 & 1.5   \\
\textbf{Cauchy Limit Phase 1} & 1.0 & 1.0 & 0.0001 & 0.0 & 0.0 & 0.0001 & 0.5  & {0.125} & 1000.0 & 0.0  & 1.5 & 1000.0\\
\textbf{Cauchy Limit Phase 2} & 2.0 & 2.0 & 0.0001 & 0.0 & 0.0 & 0.0001 & 0.75 & {0.25}  & 1000.0 & 0.0  & 1.5 & 1000.0\\
\bottomrule
\end{tabular*}
\end{table*}

{
In the results in Fig\@.~\ref{fig:plastic_strain}, we note that the Cauchy case yields at a smaller strain compared to the micropolar case.
This is due to the symmetry requirement of the Cauchy stress, so there is a non-zero $T_{21}$ component active in the Cauchy case.
Additionally, the localization of the plastic fields between the Cauchy and micropolar formulation is different.
Indeed, we observe that this non-zero $T_{21}$ component is manifested in the plastic strain maps shown along with the strain-stress curves.
Indeed, by contrasting the micropolar and Cauchy cases, we observe that the plastic strain in the micropolar case concentrates on the $x_2$-direction of the various inclusions or microstructure features, while in the Cauchy case, the plastic strain appears also in the $x_1$-direction as well.
Such difference is expected because, micropolar strains in Eq\@.~\eqref{eqn:strains} may generally be asymmetric and the micropolar balance of angular momentum in Eq\@.~\eqref{eqn:momemtum_balance} does not enforce symmetry of the stress tensor. {Lastly, we note the differences when comparing the composite responses to the responses of the matrix alone. We see an expected convergence of the curves going right to left as the volume fraction of the matrix increases, as well as a softened transition from elastic to plastic going left to right.}
}

\begin{figure*}[!ht]
    \centering
    \includegraphics[width=\textwidth]{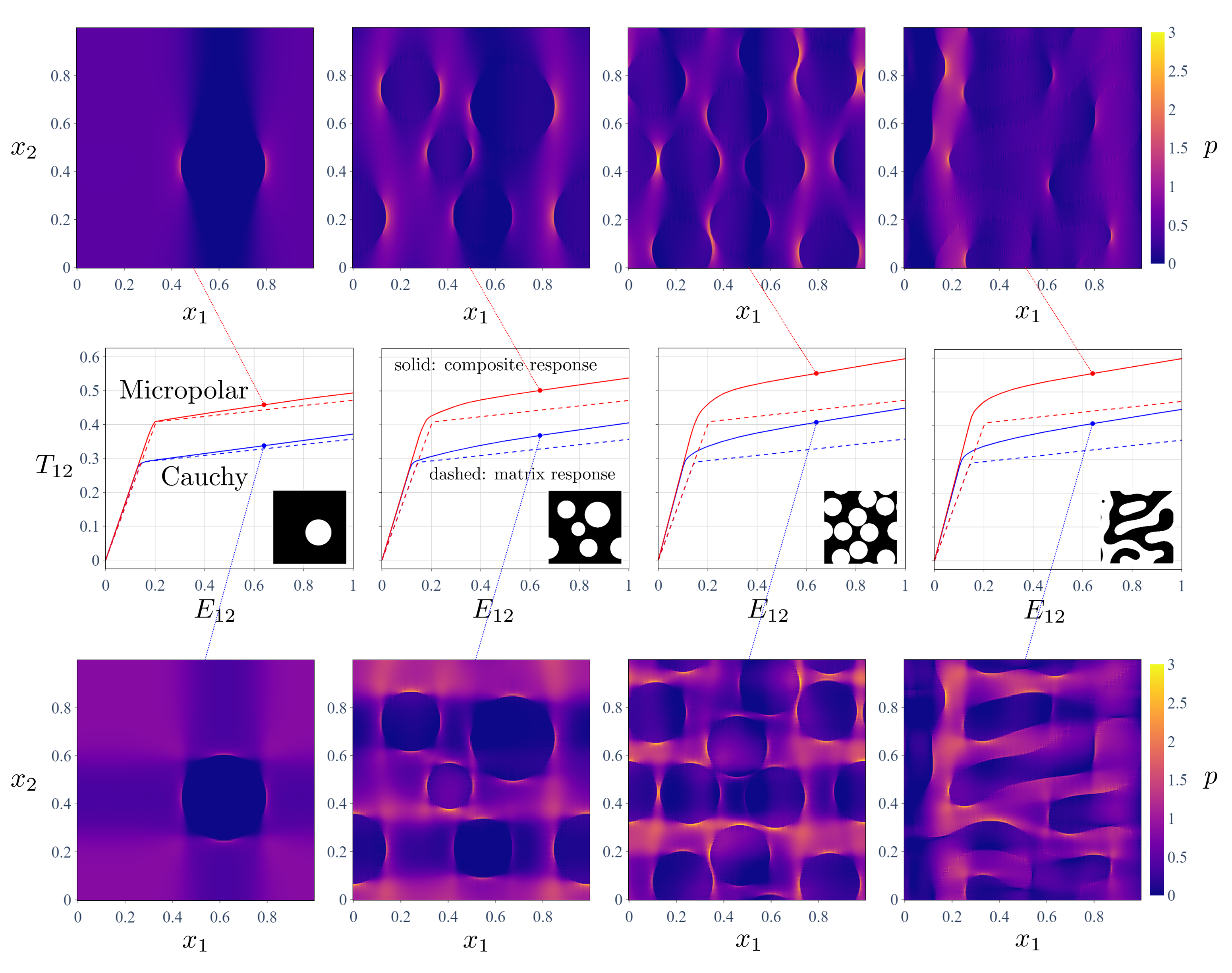}
    \caption[]{Plastic strain and averaged stress-strain curves for four geometries. Each geometry is simulated with micropolar and classical Cauchy parameters. Solid lines show the composite response, {while dashed lines show the response of the matrix (softer black material phase)}.} 
    \label{fig:plastic_strain}
\end{figure*}

\subsection{Microplastic ratcheting}
\label{sec:microratcheting}
{
Owing to the additive partitioning of the mechanical dissipation in a micropolar elastoplastic continuum (see Eq\@.~\eqref{eqn:reduced_dissipation_ineq}), even if the macro-scale stays in the elastic regime, a micropolar continuum can exhibit mechanical dissipation due to the micropolar degrees of freedom representing the underlying microstructure. 
This is notably the case in microplastic racheting, which corresponds to the progressive accumulation of plastic deformation during cyclic loading while the material is being loaded below its nominal (macroscopic) yield strength.
In metals, microplastic ratcheting finds its roots in the presence of an underlying microstructure~\cite{dingreville2010effect} which plastifies due to unfavorable orientation of its grains.
When modeled as a continuum phenomenon, microplastic ratcheting cannot be modeled with a classical Cauchy continuum if the (macroscopic) load stays in the elastic regime (i.e. there is no plastic dissipation in that case).  
}

{
In this section, we demonstrate that a micropolar elastoplastic formulation can efficiently capture this phenomenon by carefully choosing the loading path and constitutive parameters.
To that end, we required that the state of stress $t_{\text{eq}}(\mathbf{t}(\mathbf{x},t))$ is below the macro-level yield stress $t_{\text{Y}}(\mathbf{x})$ while the state of couple stress $m_{\text{eq}}(\mathbf{m}(\mathbf{x},t))$ is above the micro-level yield stress $m_{\text{Y}}(\mathbf{x})$.
The main idea here is to cyclically load the material elastically at the macro-scale, while simultaneously plastifying at the micro-scale, causing dissipation that can be ultimately observed as a macro-scale hysteresis after many loading cycles due to the coupling between the micro- and macro-scales.
}

{
To illustrate this, we chose a geometry consisting of a $50\%$ volume fraction laminate with $4\times4\times4$ voxels (see inset of Fig\@.~\ref{fig:microratcheting}), as this is one of the simplest geometries whose heterogeneity activates the couple stress $m_{kl}(\mathbf{x},t)$ even when the loading is driven only by the average macrostrain $E_{kl}(t)$.
The ratcheting loading conditions consisted of an average strain-driven loading path defined by 10 one-second cycles where $\dot{E}_{12}(t) = 1.0$ when loading and $\dot{E}_{12}(t) = -1.0$ when unloading, where we take $\Delta t = 0.01$, the number of time steps $N = 1000$, and the other components of strain rate to be zero.
The constitutive parameters were then set such that the response of the stress $t_{kl}(\mathbf{x},t)$ would stay elastic (i.e. $t_{\text{eq}}(\mathbf{t}(\mathbf{x},t)) < t_{\text{Y}}(\mathbf{x})$) and the couple stress $m_{kl}(\mathbf{x},t)$ goes plastic every cycle.
The corresponding material parameters are listed in Table~\ref{table:micropolar_phases_racheting}. 
{Here, all the values are normalized with respect to the shear modulus of phase 1, and we used a cubic side length set as $L=1$ for the geometry.}
The error threshold for the FFT-based iterative algorithm of $\varepsilon = 1\times10^{-5}$ was used with the error metric defined in Eq\@.~\eqref{eqn:err_local}. 
{Tighter error thresholds give diminishing returns in terms of accuracy, as we get a marginal error improvement at the cost of more simulation time. Fig\@.~\ref{fig:error_convergence} shows this assertion.
We use an error metric defined as
\begin{equation}
    \left\langle \left|T_{12}^{(\varepsilon)}(t) \,-\, T_{12}^{(10^{-9})}(t)\right|\right\rangle_{\text{total simulation time}}\,,
\label{eqn:err_metric_conv}
\end{equation}
where $T_{12}^{(\varepsilon)}(t)$ is the spatial average of stress $t_{12}(\mathbf{x},t)$ at time $t$ found numerically via  Alg\@.~\eqref{alg:fft} with FFT error tolerance $\varepsilon$.
We observe that the time-averaged distance from the reference solution (obtained for $\varepsilon = 1\times10^{-9}$) converges to the same level of accuracy as the reference solution approximately after $\varepsilon = 1\times10^{-5}$ (see curve with square symbol), while at the same time the computational time required to run the simulation increases nonlinearly (see curve with circular symbol).
These convergence results point to the trade-off between accuracy and efficiency of the numerical algorithm.}
}

\begin{table*}[!ht]%
\centering
\caption{Micropolar constitutive parameters of phases 1 and 2 for the ratcheting example.}
\label{table:micropolar_phases_racheting}
\begin{tabular*}{\textwidth}{@{\extracolsep\fill}lllllllllllll@{\extracolsep\fill}}%
\toprule
\textbf{Parameter} & $\lambda$ & $\mu$ & $\kappa$ & $\alpha$ & $\beta$ & $\gamma$ & $t_{\text{Y}}$ & $t_{\text{H}}$ & $m_{\text{Y}}$ & $m_{\text{H}}$ & $a_1$ & $b_1$ \\
\midrule
\textbf{Phase 1} & 1.0 & 1.0 & 1.0 & 0.0 & 1.0 & 2.0 & 2.5 & 0.0 & 0.005 & {0.0025} & 1.5 & 1.5\\
\textbf{Phase 2} & 2.0 & 2.0 & 2.0 & 0.0 & 2.0 & 4.0 & 2.5 & 0.0 & 0.005 & {0.0025} & 1.5 & 1.5\\
\bottomrule
\end{tabular*}
\end{table*}

\begin{figure*}[!ht]
    \centering
    \includegraphics[width=0.75\textwidth]{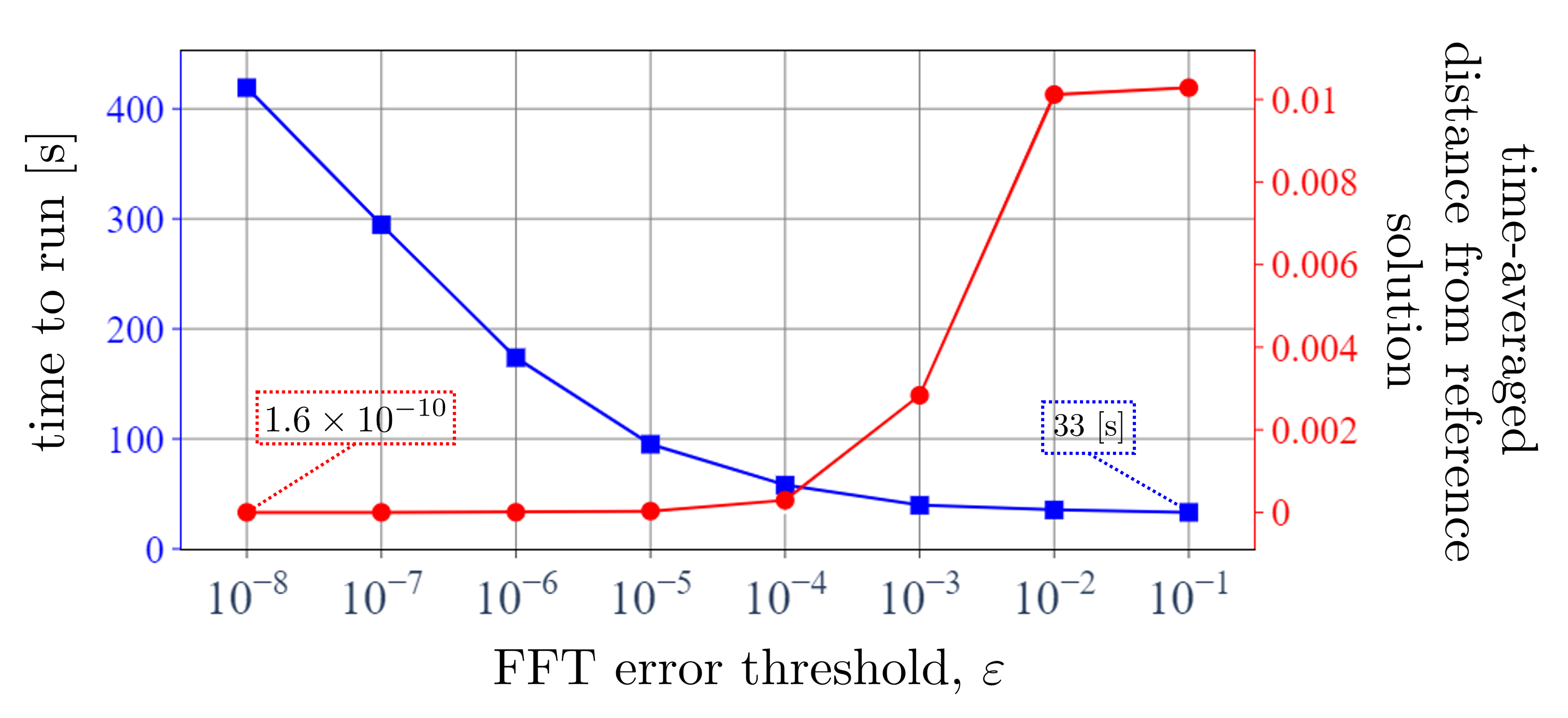}
    \caption[]{{Convergence study for the error threshold $\varepsilon$. Metric used is the time-averaged distance of $T_{12}$ from the reference solution $T_{12}$ which is taken to be the reference solution found for an error threshold $\varepsilon = 1\times10^{-9}$. The metric is defined in Eq\@.~\eqref{eqn:trial_stresses2}. The solution converges with respect to this metric after $\varepsilon = 1\times10^{-5}$, and the time the simulation takes to run increases.}} 
    \label{fig:error_convergence}
\end{figure*}

{
Fig\@.~\ref{fig:microratcheting} shows the results of cyclic loading. 
In the first row of Fig\@.~\ref{fig:microratcheting}, the resultant average stress component $T_{12}(t) := \langle t_{12}(\mathbf{x},t)\rangle$ is plotted against $E_{12}(t)$ within an inset zooming-in on the shear strain between 0.2 and 0.3;
in the second row, the average equivalent macro-level stress $T_{\text{eq}}(t) := \langle t_{\text{eq}}(\mathbf{t}(\mathbf{x},t))\rangle$ is plotted against time, and the average equivalent micro-level stress $M_{\text{eq}}(t) := \langle m_{\text{eq}}(\mathbf{m}(\mathbf{x},t))\rangle$ is plotted against time.
For $T_{\text{eq}}(t)$, the cycling remains macro-elastic, as the load stays beneath the initial yield stress value.
For $M_{\text{eq}}(t)$, the initial micro-level yield stress is plotted as a dotted line, showing that every cycle activates micro-plasticity.
Thus, under these ratcheting loading conditions, we observe that mechanical energy is dissipated, even though the behavior remains elastic at the macro-scale.
This energy dissipation can be seen at the macro-scale as an hysteresis in the $T_{12}(t)$--$E_{12}(t)$ phase space in the first row of Fig\@.~\ref{fig:microratcheting}. 
The same plot shows the solution under the same loading when the macro and micro scale are both elastic (dash-dotted line).
Notice that, in this case, there is no hysteresis as there is no energy dissipation.
}

{
If the material is cycled long enough, the above effect can cause macro-scale plasticity as well.
To show this effect, we plot the effective plastic stresses averaged over the spatial domain.
{Fig}\@.~\ref{fig:microratcheting_2} shows results for cyclic loading for the same geometry with the same loading path as those presented in Fig\@.~\ref{fig:microratcheting}, but for longer fatigue loading with 50 one-second cycles. 
The material parameters used in Fig\@.~\ref{fig:microratcheting_2}(a) are the same as the ones used in Fig\@.~\ref{fig:microratcheting}, except that the macro-scale yield stresses for materials 1 and 2 are decreased to $t_{\text{Y}}(\mathbf{x}) = 1.9~\forall~{\mathbf{x}}$, and the micro-scale hardening is set to zero $m_{\text{H}}(\mathbf{x})=0.0~\forall~{\mathbf{x}}$.
In this case, if we look at $T_{\rm eq}(t)$ in Fig\@.~\ref{fig:microratcheting_2}(a), we see that no macro-plasticity occurs, even though the average effective plastic micro strain is non-zero (see inset).
$Q(t)$ increases linearly {(see $q$ defined in Eq\@.~\eqref{eqn:trial_stresses2} with $m_{\rm H} = 0$)} and the state of couple stress $M_{\rm eq}(t)$ does not exceed the micro-level yield stress $m_{\rm Y}$.
However, when the micro-scale hardening is turned on with $m_{\text{H}}(\mathbf{x}) = {0.0025}~\forall~{\mathbf{x}}$, we observe the coupling between the micro- and the macro-level plasticity.
This case is illustrated in Fig\@.~\ref{fig:microratcheting_2}(b) where we observe that the progressive accumulation of $M_{\rm eq}(t)$ eventually leads $t_{\rm eq}(\mathbf{t}(\mathbf{x},t))$ into the plastic regime for some material point in the body (see inset where $P(t)$ is now increasing due to plasticity on the macro-scale). 
Note also that the \textit{average} equivalent macro-level stress $T_{\text{eq}}(t)$ appears not to go into the plastic regime: this is due to plasticity occurring locally and averaging out to a value that falls beneath the prescribed yield strength $t_{\text{Y}}$.
The effect of micro-plasticity causing macro-plasticity is due to the kinetic coupling intrinsic to the micropolar balance of angular momentum (see Eq\@.~\eqref{eqn:momemtum_balance}).
This behavior, which cannot be represented with the classical Cauchy model, is efficiently recreated here within the micropolar elastoplastic framework.
}
\begin{figure*}[!ht]
    \centering
    \includegraphics[width=0.75\textwidth]{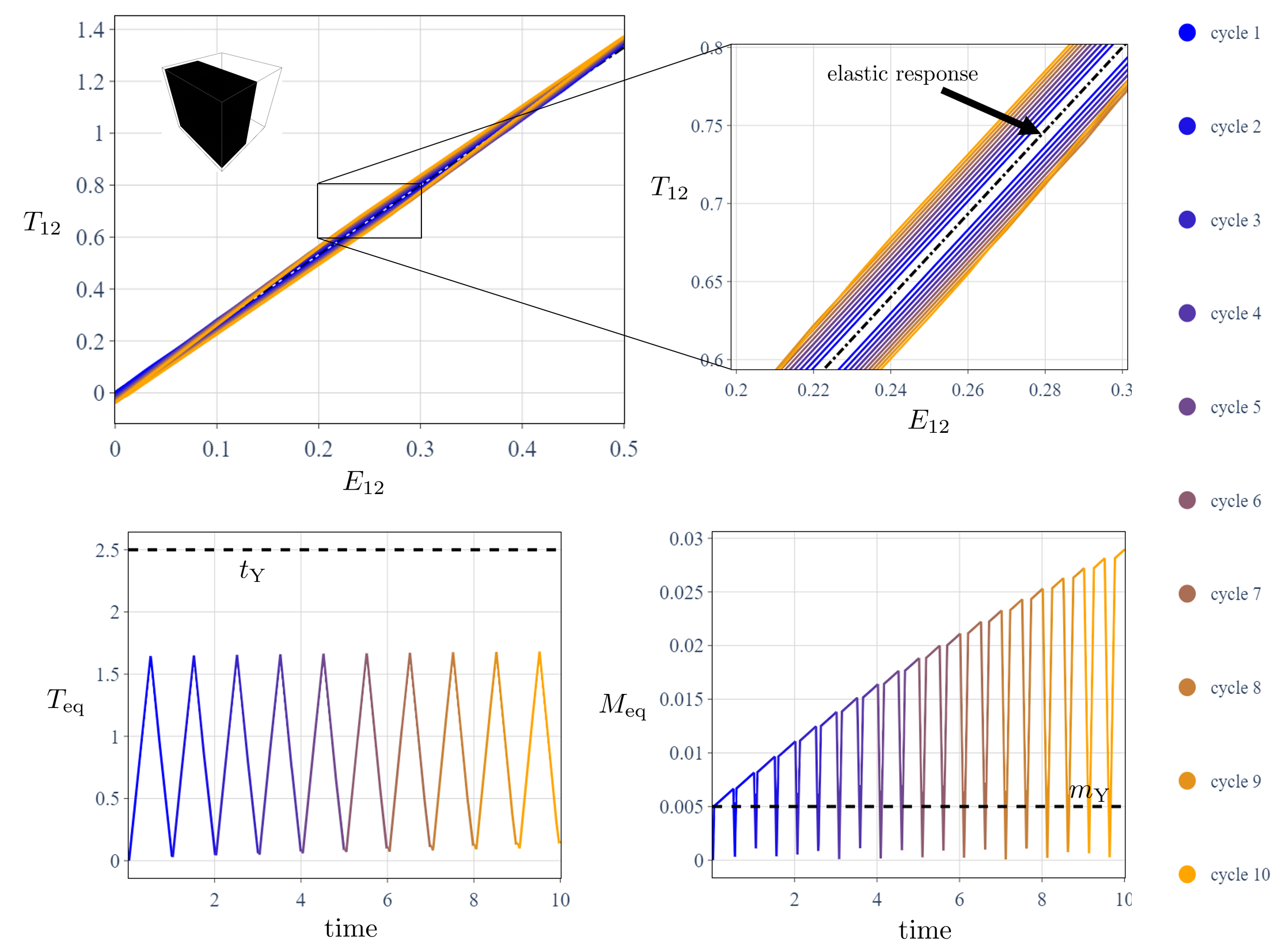}
    \caption[]{Top row: $T_{12}(t) := \langle t_{12}(\mathbf{x},t)\rangle$ as a function of $E_{12}(t)$ showing hysteresis that emerges on the macro-scale due entirely to micro-plastic dissipation. The line corresponding to the non-dissipative elastic response is shown for comparison. Bottom row: first is $T_{\text{eq}}(t) := \langle t_{\text{eq}}(\mathbf{x},t)\rangle$ as a function of time, showing that the macro-scale does not plastify; second is $M_{\text{eq}}(t) := \langle m_{\text{eq}}(\mathbf{x},t)\rangle$ as a function of time, demonstrating that the micro-scale variables are the only source of plastic dissipation in this study.} 
    \label{fig:microratcheting}
\end{figure*}
\begin{figure*}[!ht]
    \centering
    \includegraphics[width=0.75\textwidth]{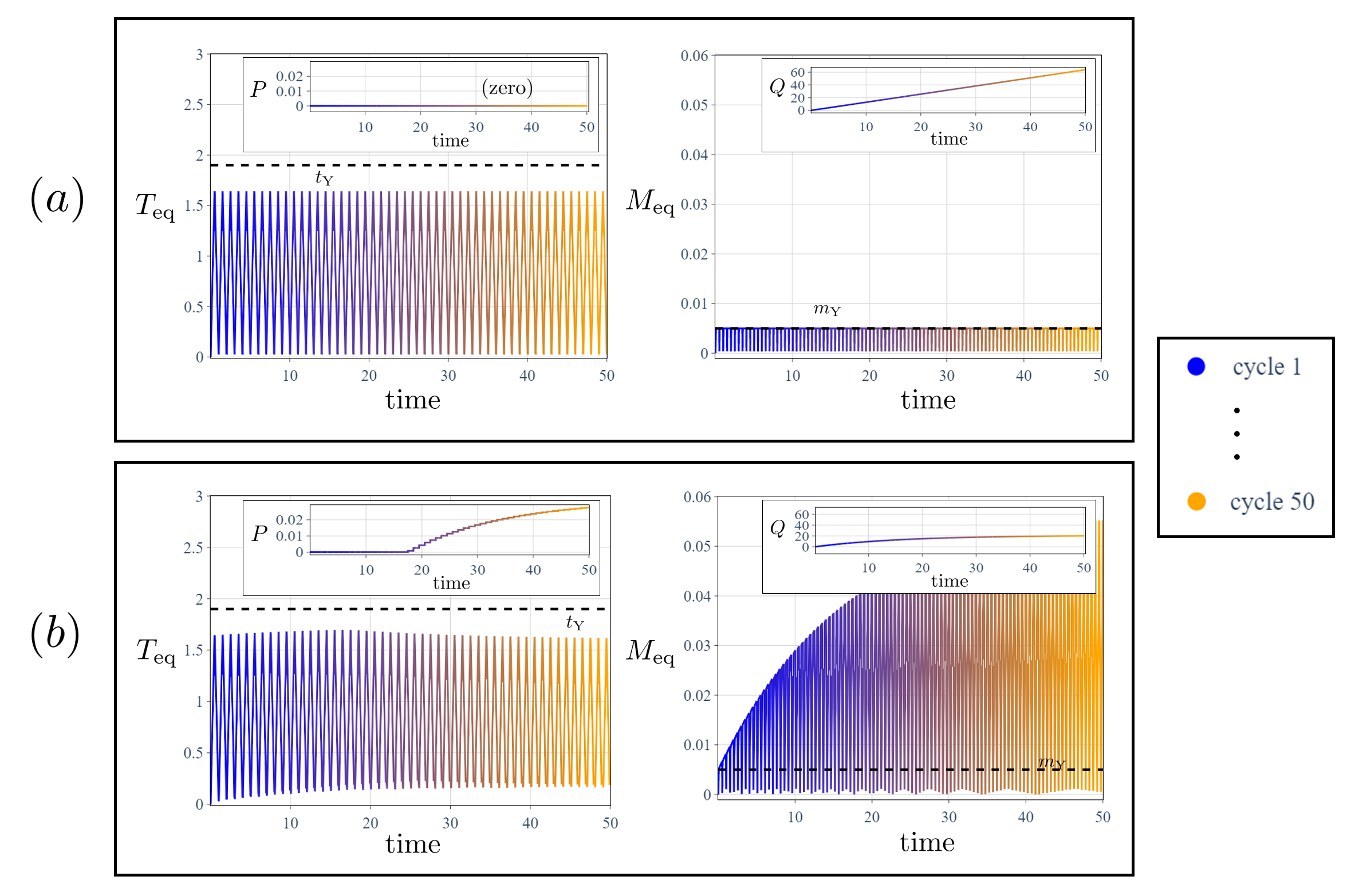}
    \caption[]{Plots of average equivalent stress on the macro- and micro-scales: $T_{\text{eq}}(t)$ and $M_{\text{eq}}(t)$, respectively. $P(t) := \left\langle p(\mathbf{x},t)\right\rangle$ and $Q(t) := \left\langle q(\mathbf{x},t)\right\rangle$ are the respective average equivalent macro- and micro-scale plastic strains. (a) Reference case for a zero micro hardening $m_{\text{H}}$. $P(t)$ is zero, so there is no macro-plasticity occurring. (b) When the micro hardening $m_{\text{H}}$ is non-zero, it is possible for macro-plastic behavior to happen. $P(t)$ becomes positive, so there is macro-plasticity occurring somewhere in the body.} 
    \label{fig:microratcheting_2}
\end{figure*}

\subsection{Length-scale effects}
\label{sec:length_scale}
{
We now turn our attention to size effects.
The micropolar model introduces length scales through the constitutive equations that control the micro-to-macro coupling.
While these length scales may be defined in various ways through dimensional analysis, here we follow \cite{russo2020thermomechanics} and define elastic and plastic length scales respectively as
\begin{equation}
    \begin{aligned}
    l^{\text{e}} ~:=~ \sqrt{\frac{\gamma}{\mu}} \,,\quad\quad l^{\text{p}} ~:=~ \sqrt{\frac{a}{b}}\,,
    \end{aligned}
    \label{eqn:ep_lengths}
\end{equation}
where $a = a_1$ in Eq\@.~\eqref{eqn:equiv_stress} and $b = b_1$ in Eq\@.~\eqref{eqn:equiv_couple_stress}. The elastic characteristic length $l^{\text{e}}$ can be identified as twice the bending characteristic length scale: see Eq\@.~(7) in Lakes \cite{lakes2016physical} and the references contained within.
The plastic characteristic length $l^{\text{p}}$, on the other hand, represents a transition length scale between the macro- and micro-level stresses since in our formulation $t_{\text{eq}} \propto \sqrt{a}$ and $m_{\text{eq}} \propto \sqrt{b}$ (see Eqs\@.~\eqref{eqn:equiv_stress} and \eqref{eqn:equiv_couple_stress}).
}

{
In this section, we study how a micropolar elastoplastic response varies when $l^{\text{e}}$ and $l^{\text{p}}$ are varied to be larger or of-the-same-order as the fixed length scale $L=1$ associated with the geometry of the microstructure.
We adopt $0.1 \leq l^{\text{e}} \leq 1.0$ and $0.1 \leq l^{\text{p}} \leq 1.0$ and generate the constitutive parameters by taking $\gamma$ and $b$ fixed and solving for $\mu$ and $a$.
Note that we do not sample internal length scales that are greater than the geometric length scale $L$, as their interpretation as \textit{microstructural} characteristic lengths breaks down.
See parameters used in Table~\ref{table:sizeeffect}.
The adopted geometry is a $16\times16\times16$ voxel representation of a matrix containing four spherical inclusions with $20\%$ volume fraction, shown in Fig\@.~\ref{fig:lengthscales}.
Arbitrarily, the average strain rates imposed (monotonically) are $\dot{E}_{32}(t) = 1$, $\dot{\Gamma}_{11}(t) = 1$, and zeros otherwise.
The time stepping parameters are $\Delta t = 0.01$ and $N = 100$. The error threshold of the FFT-based iterative algorithm is $\varepsilon = 1\times10^{-5}$ with the error metric defined in Eq\@.~\eqref{eqn:err_local}.
}

\begin{table*}[!ht]%
\centering
\caption{Micropolar constitutive parameters of phases 1 and 2 for the length scale example.}
\label{table:sizeeffect}
\begin{tabular*}{\textwidth}{@{\extracolsep\fill}lllllllllllll@{\extracolsep\fill}}%
\toprule
\textbf{Parameter} & $\lambda$ & $\mu$ & $\kappa$ & $\alpha$ & $\beta$ & $\gamma$ & $t_{\text{Y}}$ & $t_{\text{H}}$ & $m_{\text{Y}}$ & $m_{\text{H}}$ & $a$ & $b$ \\
\midrule
\textbf{Phase 1} & ${\gamma}/{(l^{\text{e}})^2}$ & ${\gamma}/{(l^{\text{e}})^2}$ & ${\gamma}/{(l^{\text{e}})^2}$ & 0.0 & ${\gamma}/{2}$ & 1.0 & 0.5 & {0.125} & 0.5 & {0.125} & $b \cdot (l^{\text{p}})^2$ & 1.5\\
\textbf{Phase 2} & ${\gamma}/{(l^{\text{e}})^2}$ & ${\gamma}/{(l^{\text{e}})^2}$ & ${\gamma}/{(l^{\text{e}})^2}$ & 0.0 & ${\gamma}/{2}$ & 2.0 & 0.75 & {0.25} & 0.75 & {0.25} & $b \cdot (l^{\text{p}})^2$ & 1.5\\
\bottomrule
\end{tabular*}
\end{table*}

{
The main point of Fig\@.~\ref{fig:lengthscales} is to show that the micropolar elastoplastic model at-hand is sensitive to different choices of the internal lengths defined in Eq\@.~\eqref{eqn:ep_lengths}. To demonstrate this, due to monotonic loading over time, we take the last value of averaged stress $T_{32}(t_{\text{last}}) := \langle t_{32}(\mathbf{x},t_{\text{last}}) \rangle$ and averaged couple stress $M_{11}(t_{\text{last}}) := \langle m_{11}(\mathbf{x},t_{\text{last}}) \rangle$ and plot them on a heatmap as a function of both length scales. 
}

{
Since we take the micropolar parameters of the problem to be constant across all length scales, $M_{11}(t_{\text{last}})$ in Fig\@.~\ref{fig:lengthscales} shows as a constant heatmap. 
On the other hand, the heatmap for $T_{32}(t_{\text{last}})$ in Fig\@.~\ref{fig:lengthscales} depicts a gradient of color representing the variable's sensitivity to the elastic and plastic length scales.
As seen from Eq\@.~\eqref{eqn:ep_lengths} and Table~\ref{table:sizeeffect}, the macro-scale response is elastically stiffer when $l^{\text{e}}$ is small compared to $L$, and more compliant when $l^{\text{e}}$ is of the same order. 
Plastically, the material has more difficulty yielding when $l^{\text{p}}$ is small, and an easier time yielding when $l^{\text{p}}$ is of the same order.
Starting from the bottom-left part of the $T_{32}(t_{\text{last}})$ heatmap in Fig\@.~\ref{fig:lengthscales}, there is a region where both length scales are much less than the geometric length scale: $l^{\text{e}} \sim l^{\text{p}} \ll L$.
The plot of the macro-response over time shows high elastic stiffness leading to almost immediate yielding and hardening.
We see also in this region that $T_{32}(t_{\text{last}}) \approx 14 \gg 0.5 \approx M_{11}(t_{\text{last}})$.
This means that in this region, the classical Cauchy model takes over and governs the response of the material as the microstructural effects become negligible. 
Looking at the micropolar balance laws in Eq\@.~\eqref{eqn:momemtum_balance}, if the divergence of the couple stress $m_{kl,k}$ and the body couple $l_l$ are negligible compared to $t_{kl}$, then classical balance of angular momentum holds, forcing the stress tensor $t_{kl}$ to be symmetric.
Moving up the plot to the top-left part, we have a region that represents when the elastic length is of the same order, but the plastic length is much less than the geometric length: $l^{\text{p}} \ll l^{\text{e}} \sim  L$.
We see in the plot of the macro-response over time that this combination leads to a more compliant elastic response that never becomes plastic in the simulation window.
Comparing the magnitude of the response, $T_{32}(t_{\text{last}}) \approx 4 > 0.5 \approx M_{11}(t_{\text{last}})$, we start to see that the microstructural effects are non-negligible.
Lastly, when we move to the right side of the plot where the plastic length scale is of the same order compared to the geometric length ($l^{\text{p}} \sim L$), for all choices of the elastic length scale there is a region where a fully-coupled micropolar response across scales takes place, since $M_{11}(t_{\text{last}}) \sim T_{32}(t_{\text{last}})$. Looking at the macro-responses over time, note that: 1) when the elastic length is small, there is a stiffer response, 2) when the elastic length is bigger, there is a more compliant response.
}

\begin{figure*}[!ht]
    \centering
    \includegraphics[width=\textwidth]{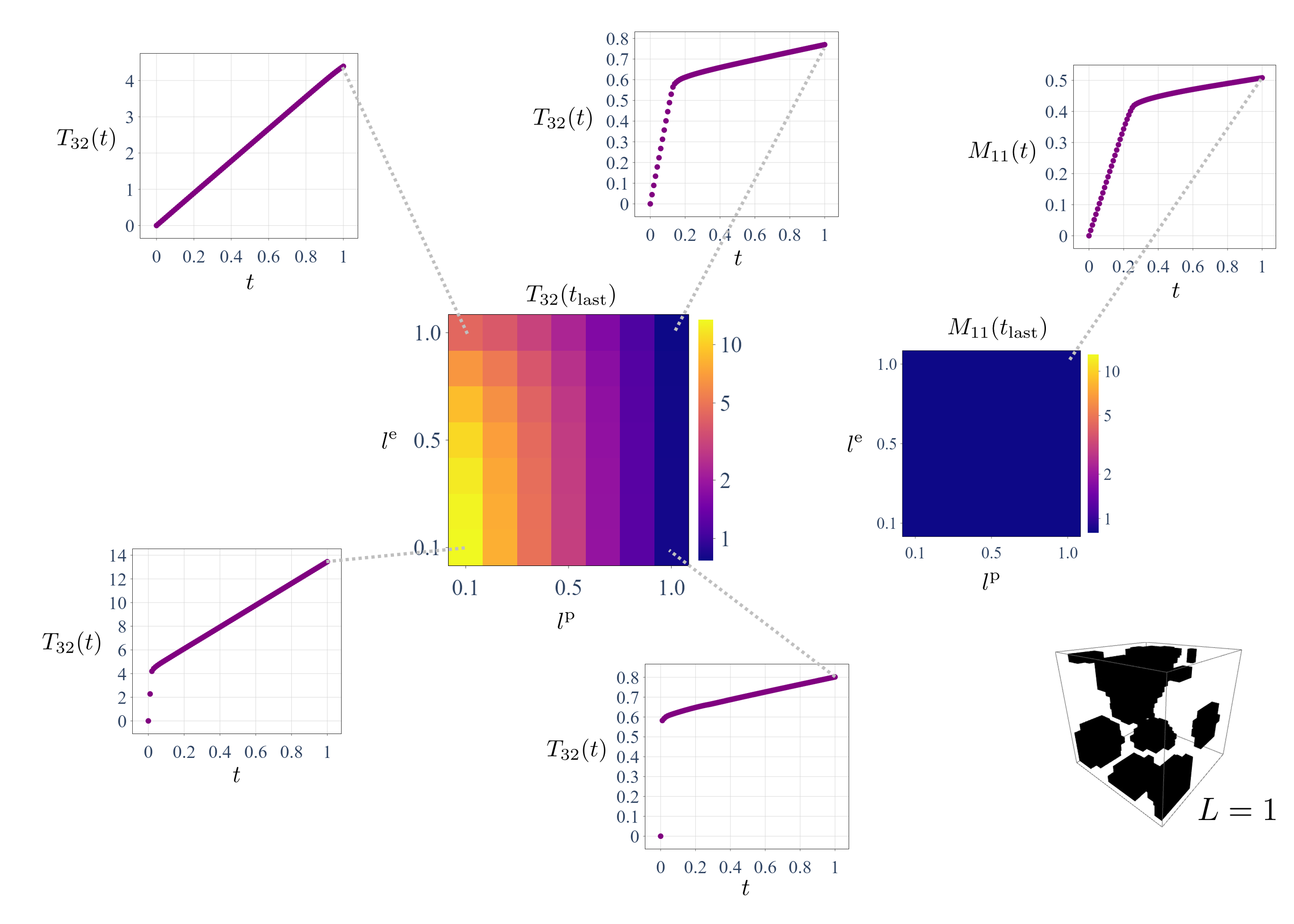}
    \caption[]{Varying classical constitutive parameters $\mu$, $a$ and keeping micropolar constitutive parameters $\gamma$ and $b$ fixed to scan the full range of elastic $l^{\text{e}}$ and plastic $l^{\text{p}}$ length scales defined in the figure and in Eq\@.~\eqref{eqn:ep_lengths}. Geometry used is four spherical inclusions inside a matrix with side length $L = 1$. The limit $l^{\text{e}} \sim l^{\text{p}} \sim L$ (as well as $l^{\text{e}} \ll l^{\text{p}} \sim L$) is where the macro and micro scales are fully coupled in the micropolar model. When $l^{\text{e}} \sim l^{\text{p}} \ll L$, the micropolar model recovers the classical Cauchy model.} 
    \label{fig:lengthscales}
\end{figure*}

\subsection{{Time scalability and fixed-point iterations per time step}} 
\label{sec:complexity}
This section presents a study of space resolution and computing time of Alg\@.~\ref{alg:fft}.
For consistency, {simulations were performed} on a single Intel Core i7-3520M CPU with 12GB of RAM installed. 
For every resolution, we ran Alg\@.~\ref{alg:fft} ten individual times, collecting the time it took to finish and the amount of storage the allocations used.
We then averaged the times and storage amounts over the ten runs and plot the results in Fig\@.~\ref{fig:TC}.
The geometry used in this study is the 50\% volume fraction laminate.
The material parameters chosen are listed in Table~\ref{table:comput}.
The cubic side length is set as $L=1$.
The average strain rates imposed are $\dot{E}_{13}(t) = 1$, $\dot{\Gamma}_{32}(t) = 1$, and zero for all other components.
For time stepping, $\Delta t = 0.01$ and $N=100$, and for the threshold error of the FFT-based algorithm we use $\varepsilon=1\times10^{-5}$ with Eq\@.~\eqref{eqn:err_local} as the error metric.
The number of voxels for the six resolutions considered are $1$, $8$, $64$, $512$, $4096$, and $32768$. 
Fig\@.~\ref{fig:TC} shows that the calculation time as a complexity of $n\log(n)$ instead of  $n^{2}$ as it would be the case for a finite-element implementation.
Our implementation pre-calculates the Green's tensors needed for the FFT iterations, requiring less time, but more memory than if the Green's tensors were computed fresh at each iteration. Note that in Fig\@.~\ref{fig:TC} the storage unit of mebibyte [MiB] ($1024^2$ bytes) is used, not megabyte [MB] ($1000^2$ bytes).
\begin{figure*}[!ht]
    \centering
    \includegraphics[width=0.65\textwidth]{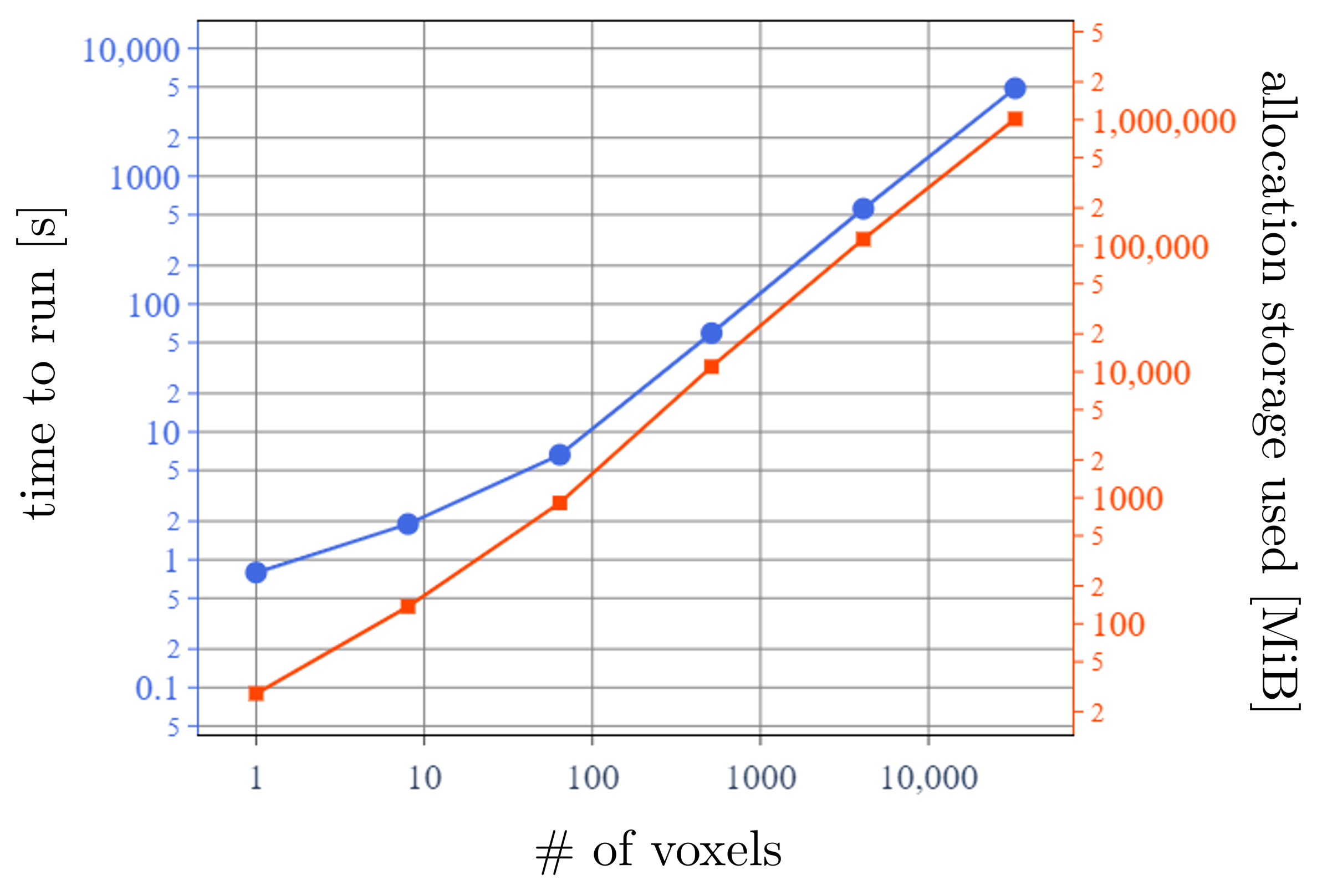}
    \caption[]{Computing time and memory storage as a function of spacial resolution for Alg\@.~\ref{alg:fft}.} 
    \label{fig:TC}
\end{figure*}

\begin{table*}[!ht]%
\centering
\caption{Micropolar constitutive parameters for phases 1 and 2 for the space time complexity study.}
\label{table:comput}
\begin{tabular*}{\textwidth}{@{\extracolsep\fill}lllllllllllll@{\extracolsep\fill}}%
\toprule
\textbf{Parameter} & $\lambda$ & $\mu$ & $\kappa$ & $\alpha$ & $\beta$ & $\gamma$ & $t_{\text{Y}}$ & $t_{\text{H}}$ & $m_{\text{Y}}$ & $m_{\text{H}}$ & $a_1$ & $b_1$ \\
\midrule
\textbf{Phase 1} & 1.0 & 1.0 & 1.0 & 0.0 & 0.5 & 1.0 & 0.5  & {0.125} & 0.5  & {0.125} & 1.5 & 1.5\\
\textbf{Phase 2} & 2.0 & 2.0 & 2.0 & 0.0 & 1.0 & 2.0 & 0.75 & {0.25}  & 0.75 & {0.25}  & 1.5 & 1.5\\
\bottomrule
\end{tabular*}
\end{table*}

{Under the same prescribed strain rates and time stepping parameters, we also studied how varying the nonlinear parameters affects simulation time via the number of fixed-point iterations needed per step.
Here we used the 512-voxel 50\% volume fraction laminate geometry with cubic side length $L=1$.
The threshold error of the FFT-based algorithm used for these simulation was set to $\varepsilon=1\times10^{-8}$ with Eq\@.~\eqref{eqn:err_local} as the error metric.
The range of nonlinear material parameters used for this set of simulations is provided in Table~\ref{table:iter_step}, where we varied the hardening moduli of phase 2 as $t_{\rm H}^2 = x,~m_{\rm H}^2 = x$ with $x \in \{0,0.001,0.005,0.01,0.05,0.1\}$ (value of the hardening moduli of phase 1 are kept constant to a value of $t_{\rm H}^1 = 0.1,~m_{\rm H}^1 = 0.1$).
Plotted in Fig\@.~\ref{fig:iters_vs_steps}, are six curves showing fixed-point iterations as a function of the time step as we varied $x$.
We observe that, as $x\rightarrow 0$ (i.e. as the hardening difference between phase 1 and phase 2 becomes large), the simulation time increases (assuming a fixed time per fixed-point iteration).
This trend indicates that the simulation time of Alg\@.~\ref{alg:fft} depends on the nonlinear hardening parameters, since they control the relative tangent
stiffness when the full plastic regime is reached.}

\begin{figure*}[!ht]
    \centering
    \includegraphics[width=0.75\textwidth]{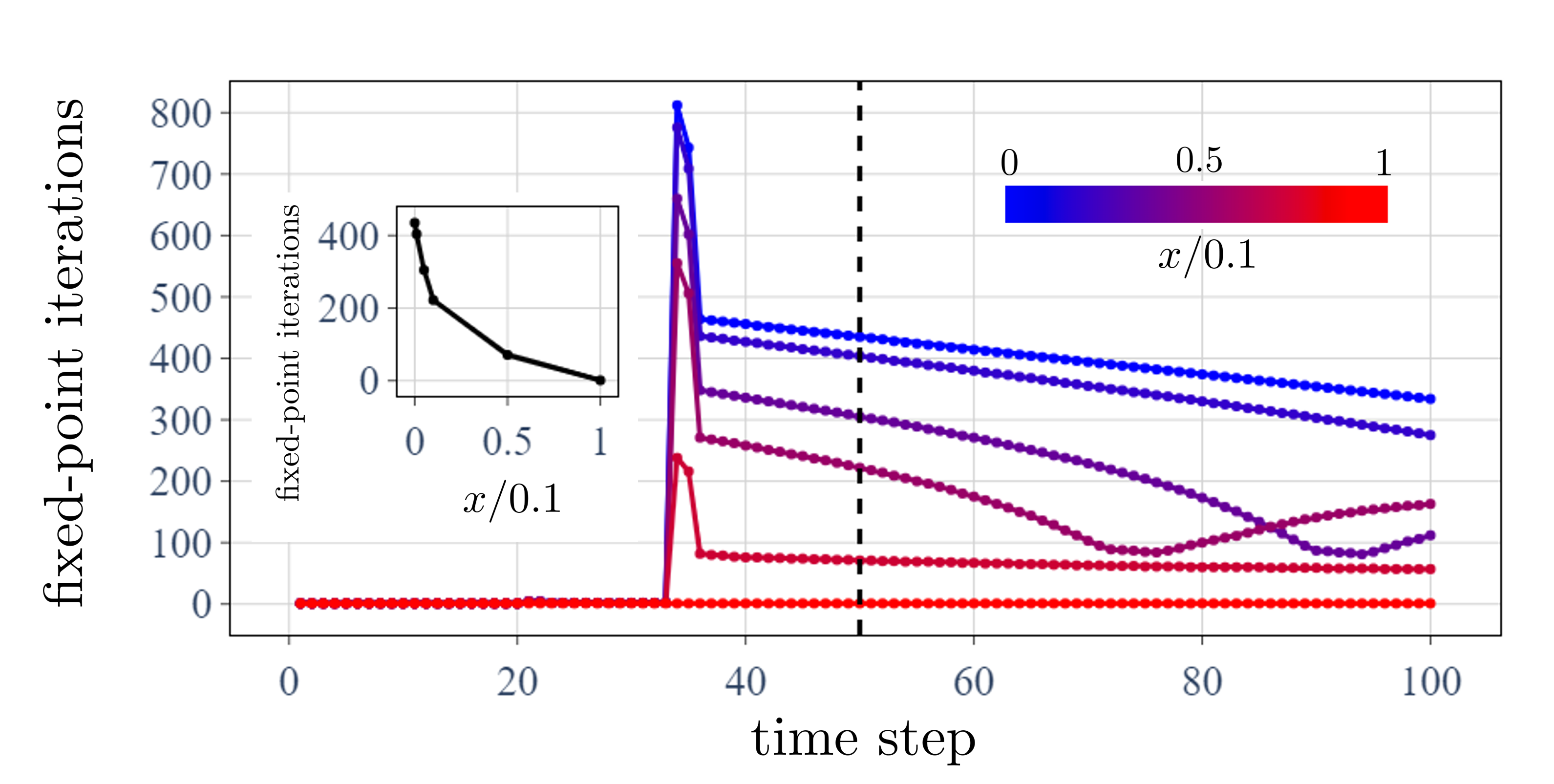}
    \caption[]{{Fixed-point iterations vs time step across different hardening modulus contrasts. Iterations increase as the hardening moduli (both $t_{\text{H}}^2$ and $m_{\text{H}}^2$) go to zero with phase 1 hardenings fixed. Simulation time depends strongly on the nonlinear parameters as well as the elastic heterogeneity.}} 
    \label{fig:iters_vs_steps}
\end{figure*}

\begin{table*}[!ht]%
\centering
\caption{{Micropolar constitutive parameters for phases 1 and 2 for the iterations per time step study.}}
\label{table:iter_step}
\begin{tabular*}{\textwidth}{@{\extracolsep\fill}lllllllllllll@{\extracolsep\fill}}%
\toprule
\textbf{Parameter} & $\lambda$ & $\mu$ & $\kappa$ & $\alpha$ & $\beta$ & $\gamma$ & $t_{\text{Y}}$ & $t_{\text{H}}$ & $m_{\text{Y}}$ & $m_{\text{H}}$ & $a_1$ & $b_1$ \\
\midrule
\textbf{Phase 1} & 1.0 & 1.0 & 1.0 & 0.0 & 0.5 & 0.1 & 0.5  & 0.1 & 0.5 & 0.1 & 1.5 & 1.5\\
\textbf{Phase 2} & 1.0 & 1.0 & 1.0 & 0.0 & 0.5 & 1.0 & 0.5 & $x$  & 0.5 & $x$  & 1.5 & 1.5\\
\bottomrule
\end{tabular*}
\end{table*}


\section{Conclusions}
In this paper, the numerical method presented previously in \cite{francis2024fast} is extended to micropolar elastoplasticity following Moulinec and Suquet \cite{Moulinec1998} for classical Cauchy elastoplasticity.
To solve the micropolar elastoplastic constitutive equations, we derived a closed-form radial-return mapping algorithm.
Our implementation was verified via an extension of the method of manufactured solutions, called MNMS-II from \cite{chen2017proposal}, and a convergence study.
We showed that using the micropolar model is possible to capture effects such as mechanical dissipation entirely due to the microstructure, and scale/size-dependent responses.
The proposed algorithm converges as long as the contrast in material properties is relatively low. For high- or infinite-contrast materials, an extension of the micropolar Augmented Lagrangian scheme presented in \cite{francis2024fast} to nonlinear constitutive behavior following \cite{Michel2000,Michel2001,lebensohn2012elasto,Upadhyay2016} can be done and will be reported in a future contribution.

\bmsection*{Acknowledgments}
The authors wish to thank Dr\@.~Dongil Shin from Sandia National Laboratories for helpful feedback and suggestions, and Hooman Dadras from the University of Colorado Boulder for valuable discussions about plasticity. {The authors would also like to thank the anonymous reviewer for the constructive feedback and suggestions to improve the overall clarity and quality of the manuscript.}

\bmsection*{Financial disclosure}
NMF received complementary financial support from the University of Colorado Boulder through FP's startup fund and a College of Engineering and Applied Science--Sandia National Laboratories (CEAS-SNL) partnership fund.
This work utilized the Alpine high performance computing resource at the University of Colorado Boulder. Alpine is jointly funded by the University of Colorado Boulder, the University of Colorado Anschutz, Colorado State University, and the National Science Foundation (award 2201538).
RAL acknowledges support from the Advanced Engineering Materials program at Los Alamos
National Laboratory. 
This work is supported in part by the Center for Integrated Nanotechnologies, an Office of Science user facility operated for the U\@.S\@.~Department of Energy.
This article has been authored by an employee of National Technology \& Engineering Solutions of Sandia, LLC under Contract No\@.~DE-NA0003525 with the U\@.S\@.~Department of Energy (DOE).
The employee owns all right, title, and interest in and to the article and is solely responsible for its contents. The United States Government retains and the publisher, by accepting the article for publication, acknowledges that the United States Government retains a non-exclusive, paid-up, irrevocable, world-wide license to publish or reproduce the published form of this article or allow others to do so, for United States Government purposes.
The DOE will provide public access to these results of federally sponsored research in accordance with the DOE Public Access Plan https://www.energy.gov/downloads/doe-public-access-plan.

\bmsection*{Conflict of interest}

The authors declare no potential conflict of interests.

\bibliography{micropolar_main}

\begin{thebibliography}{10}
\providecommand \doibase [0]{http://dx.doi.org/}%

\bibitem{Cosserat1909}
Cosserat EMP, Cosserat F. {\it Th{\'e}orie des Corps D{\'e}formables}.
\newblock A. Hermann et fils, 1909.

\bibitem{suhubi1964nonlinear}
Suhubi ES, Eringen AC. Nonlinear theory of micro-elastic solids--{II}. {\it Int J Eng Sci.} 1964\string;2(4)\string:389--404.
\newblock \url{https://doi.org/10.1016/0020-7225(64)90017-5}.

\bibitem{eringen1966linear}
Eringen AC. Linear theory of micropolar elasticity. {\it J Math Mech.} 1966\string;15(6)\string:909--923.
\newblock \url{https://www.jstor.org/stable/24901442}.

\bibitem{dingreville2005surface}
Dingreville R, Qu J, Cherkaoui M. Surface free energy and its effect on the elastic behavior of nano-sized particles, wires and films. {\it J Mech Phys Solids.} 2005\string;53(8)\string:1827--1854.
\newblock \url{https://doi.org/10.1016/j.jmps.2005.02.012}.

\bibitem{forest2000cosserat}
Forest S, Barbe F, Cailletaud G. Cosserat modelling of size effects in the mechanical behaviour of polycrystals and multi-phase materials. {\it Int J Solids Struct.} 2000\string;37(46-47)\string:7105--7126.
\newblock \url{https://doi.org/10.1016/S0020-7683(99)00330-3}.

\bibitem{mayeur2011dislocation}
Mayeur JR, McDowell DL, Bammann DJ. Dislocation-based micropolar single crystal plasticity: {C}omparison of multi- and single criterion theories. {\it J Mech Phys Solids.} 2011\string;59(2)\string:398--422.
\newblock \url{https://doi.org/10.1016/j.jmps.2010.09.013}.

\bibitem{yoder2018size}
Yoder M, Thompson L, Summers J. Size effects in lattice structures and a comparison to micropolar elasticity. {\it Int J Solids Struct.} 2018\string;143\string:245--261.
\newblock \url{https://doi.org/10.1016/j.ijsolstr.2018.03.013}.

\bibitem{dos2012construction}
Dos~Reis F, Ganghoffer JF. Construction of micropolar continua from the asymptotic homogenization of beam lattices. {\it Comput Struct.} 2012\string;112\string:354--363.
\newblock \url{https://doi.org/10.1016/j.compstruc.2012.08.006}.

\bibitem{spadoni2012elasto}
Spadoni A, Ruzzene M. Elasto-static micropolar behavior of a chiral auxetic lattice. {\it J Mech Phys Solids.} 2012\string;60(1)\string:156--171.
\newblock \url{https://doi.org/10.1016/j.jmps.2011.09.012}.

\bibitem{dingreville2014wave}
Dingreville R, Robbins J, Voth T. Wave propagation and dispersion in elasto-plastic microstructured materials. {\it Int. J. Solids Struct..} 2014\string;51(11-12)\string:2226--2237.
\newblock \url{https://doi.org/10.1016/j.ijsolstr.2014.02.030}.

\bibitem{Alberdi2021}
Alberdi R, Robbins J, Walsh T, Dingreville R. Exploring wave propagation in heterogeneous metastructures using the relaxed micromorphic model. {\it J. Mech. Phys. Solids.} 2021\string;155\string:104540.
\newblock \url{https://doi.org/10.1016/j.jmps.2021.104540}.

\bibitem{Neuner2020}
Neuner M, Gamnitzer P, Hofstetter G. A {3D} gradient-enhanced micropolar damage-plasticity approach for modeling quasi-brittle failure of cohesive-frictional materials. {\it Comput Struct.} 2020\string;239\string:106332.
\newblock \url{https://doi.org/10.1016/j.compstruc.2020.106332}.

\bibitem{Moulinec1998}
Moulinec H, Suquet P. A numerical method for computing the overall response of nonlinear composites with complex microstructure. {\it Comput Methods Appl Mech Eng.} 1998\string;157(1-2)\string:69--94.
\newblock \url{https://doi.org/10.1016/S0045-7825(97)00218-1}.

\bibitem{cooley1965algorithm}
Cooley JW, Tukey JW. An algorithm for the machine calculation of complex {F}ourier series. {\it Math Comput.} 1965\string;19(90)\string:297--301.
\newblock \url{https://doi.org/10.2307/2003354}.

\bibitem{frigo2005design}
Frigo M, Johnson SG. The {D}esign and {I}mplementation of {FFTW3}. {\it Proc IEEE.} 2005\string;93(2)\string:216--231.
\newblock \url{https://doi.org/10.1109/jproc.2004.840301}.

\bibitem{Zecevic2022}
Zecevic M, Lebensohn RA, Capolungo L. New large-strain {FFT}-based formulation and its application to model strain localization in nano-metallic laminates and other strongly anisotropic crystalline materials. {\it Mech Mater.} 2022\string;166\string:104208.
\newblock \url{https://doi.org/10.1016/j.mechmat.2021.104208}.

\bibitem{Berbenni2014}
Berbenni S, Taupin V, Djaka KS, Fressengeas C. A numerical spectral approach for solving elasto-static field dislocation and g-disclination mechanics. {\it Int J Solids Struct.} 2014\string;51(23-24)\string:4157--4175.
\newblock \url{https://doi.org/10.1016/j.ijsolstr.2014.08.009}.

\bibitem{djaka2020fft}
Djaka KS, Berbenni S, Taupin V, Lebensohn RA. A {FFT}-based numerical implementation of mesoscale field dislocation mechanics: application to two-phase laminates. {\it Int J Solids Struct.} 2020\string;184\string:136--152.
\newblock \url{https://doi.org/10.1016/j.ijsolstr.2018.12.027}.

\bibitem{Lebensohn2016}
Lebensohn RA, Needleman A. Numerical implementation of non-local polycrystal plasticity using fast {Fourier} transforms. {\it J Mech Phys Solids.} 2016\string;97\string:333--351.
\newblock \url{https://doi.org/10.1016/j.jmps.2016.03.023}.

\bibitem{Upadhyay2016}
Upadhyay MV, Capolungo L, Taupin V, Fressengeas C, Lebensohn RA. A higher order elasto-viscoplastic model using fast {Fourier} transforms: {E}ffects of lattice curvatures on mechanical response of nanocrystalline metals. {\it Int J Plast.} 2016\string;83\string:126--152.
\newblock \url{https://doi.org/10.1016/j.ijplas.2016.04.007}.

\bibitem{Segurado2021}
Segurado J, Lebensohn RA. An {FFT}-based approach for {Bloch} wave analysis: application to polycrystals. {\it Comput Mech.} 2021\string;68(5)\string:981--1001.
\newblock \url{https://doi.org/10.1007/s00466-021-02055-9}.

\bibitem{francis2024fast}
Francis NM, Pourahmadian F, Lebensohn RA, Dingreville R. A fast {F}ourier transform-based solver for elastic micropolar composites. {\it Comput Methods Appl Mech Eng.} 2024\string;418\string:116510.
\newblock \url{https://doi.org/10.1016/j.cma.2023.116510}.

\bibitem{grimm2021fft}
Grimm-Strele H, Kabel M. {FFT}-based homogenization with mixed uniform boundary conditions. {\it Int J Numer Methods Eng.} 2021\string;122(23)\string:7241--7265.
\newblock \url{https://doi.org/10.1002/nme.6830}.

\bibitem{risthaus2024fft}
Risthaus L, Schneider M. {FFT}-based computational micromechanics with {D}irichlet boundary conditions on the rotated staggered grid. {\it Int J Numer Methods Eng.} 2024\string:e7569.
\newblock \url{https://doi.org/10.1002/nme.7569}.

\bibitem{risthaus2024imposing}
Risthaus L, Schneider M. Imposing different boundary conditions for thermal computational homogenization problems with {FFT}-and tensor-train-based {G}reen's operator methods. {\it Int J Numer Methods Eng.} 2024\string;125(7)\string:e7423.
\newblock \url{https://doi.org/10.1002/nme.7423}.

\bibitem{russo2020thermomechanics}
Russo R, Forest S, Girot~Mata FA. Thermomechanics of {C}osserat medium: modeling adiabatic shear bands in metals. {\it Contin Mech Thermodyn.} 2020\string:1--20.
\newblock \url{https://doi.org/10.1007/s00161-020-00930-z}.

\bibitem{forest2003elastoviscoplastic}
Forest S, Sievert R. Elastoviscoplastic constitutive frameworks for generalized continua. {\it Acta Mech.} 2003\string;160(1)\string:71--111.
\newblock \url{https://doi.org/10.1007/s00707-002-0975-0}.

\bibitem{de1993generalisation}
{de}~Borst R. A generalisation of ${J}_2$-flow theory for polar continua. {\it Comput Methods Appl Mech Eng.} 1993\string;103(3)\string:347--362.
\newblock \url{https://doi.org/10.1016/0045-7825(93)90127-J}.

\bibitem{steinmann1991localization}
Steinmann P, Willam K. Localization within the {F}ramework of {M}icropolar {E}lasto-{P}lasticity. In:  Br\"uller OS, Mannl V, Najar J. \kern-2pt, eds. {\it Advances in Continuum Mechanics}, , Berlin, Heidelberg: Springer,  1991\string:296--313.
\newblock \url{https://doi.org/10.1007/978-3-642-48890-0_24}.

\bibitem{simo2006computational}
Simo JC, Hughes TJR. {\it Computational {I}nelasticity}. 7.
\newblock Springer Science \& Business Media, 2006.
\newblock \url{https://doi.org/10.1007/b98904}.

\bibitem{eringen1968theory}
Eringen AC. Theory of micropolar elasticity. In:  Liebowtiz H. \kern-2pt, ed. {\it Fracture: An Advanced Treatise}, , Academic Press,  1968\string:621--729.

\bibitem{Eringen1999}
Eringen AC. {\it Microcontinuum Field Theories: {I. Foundations and Solids}}.
\newblock Springer Science \& Business Media, 2012.
\newblock \url{https://doi.org/10.1007/978-1-4612-0555-5}.

\bibitem{Eringen1962}
Eringen AC. {\it Nonlinear {T}heory of {C}ontinuous {M}edia}.
\newblock McGraw-Hill, 1962.

\bibitem{nowacki1974linear}
Nowacki W. The {L}inear {T}heory of {M}icropolar {E}lasticity. In: Springer.  1974\string:1--43.
\newblock \url{https://link.springer.com/content/pdf/10.1007/978-3-7091-2920-3.pdf#page=9}.

\bibitem{ferrier2023posteriori}
Ferrier R, Bellis C. A posteriori error estimations and convergence criteria in fast {F}ourier transform-based computational homogenization. {\it Int J Numer Methods Eng.} 2023\string;124(4)\string:834--863.
\newblock \url{https://doi.org/10.1002/nme.7145}.

\bibitem{Stewart2020}
Stewart JA, Dingreville R. Microstructure morphology and concentration modulation of nanocomposite thin-films during simulated physical vapor deposition. {\it Acta Mater.} 2020\string;188\string:181--191.
\newblock \url{https://doi.org/10.1016/j.actamat.2020.02.011}.

\bibitem{dingreville2020benchmark}
Dingreville R, Stewart JA, Chen EY, Monti JM. Benchmark problems for the {Mesoscale Multiphysics Phase Field Simulator} ({MEMPHIS}). Tech. Rep. SAND2020-12852, Sandia National Laboratories (SNL-NM); Albuquerque, NM (United States):   2020.
\newblock \url{https://doi.org/10.2172/1615889}.

\bibitem{dingreville2010effect}
Dingreville R, Battaile CC, Brewer LN, Holm EA, Boyce BL. The effect of microstructural representation on simulations of microplastic ratcheting. {\it Int J Plast.} 2010\string;26(5)\string:617--633.
\newblock \url{https://doi.org/10.1016/j.ijplas.2009.09.004}.

\bibitem{lakes2016physical}
Lakes R. Physical meaning of elastic constants in {C}osserat, void, and microstretch elasticity. {\it J Mech Mater Struct.} 2016\string;11(3)\string:217--229.
\newblock \url{https://doi.org/10.2140/jomms.2016.11.217}.

\bibitem{chen2017proposal}
Chen J, Hori M, O-tani H, Oishi S, Fujita K, Motoyama H. Proposal of {M}ethod of {N}umerically {M}anufactured {S}olutions for {C}ode {V}erification of {E}lasto-{P}lastic {P}roblems. {\it J Jpn Soc Civil Eng, Ser A2 (Appl Mech).} 2017\string;73(2)\string:I\_165--I\_175.
\newblock \url{https://doi.org/10.2208/jscejam.73.I_165}.

\bibitem{Michel2000}
Michel JC, Moulinec H, Suquet P. A {C}omputational {M}ethod {B}ased on {A}ugmented {L}agrangians and {F}ast {F}ourier {T}ransforms for {C}omposites with {H}igh {C}ontrast. {\it CMES--Comp Model Eng.} 2000\string;1(2)\string:79--88.
\newblock \url{https://doi.org/10.3970/cmes.2000.001.239}.

\bibitem{Michel2001}
Michel JC, Moulinec H, Suquet P. A computational scheme for linear and non-linear composites with arbitrary phase contrast. {\it Int J Numer Methods Eng.} 2001\string;52(1-2)\string:139--160.
\newblock \url{https://doi.org/10.1002/nme.275}.

\bibitem{lebensohn2012elasto}
Lebensohn RA, Kanjarla AK, Eisenlohr P. An elasto-viscoplastic formulation based on fast {F}ourier transforms for the prediction of micromechanical fields in polycrystalline materials. {\it Int J Plast.} 2012\string;32\string:59--69.
\newblock \url{https://doi.org/10.1016/j.ijplas.2011.12.005}.

\bibitem{schwer2007overview}
Schwer LE. An overview of the {PTC} 60/{V}\&{V} 10: guide for verification and validation in computational solid mechanics: Transmitted by {L}.{E}. {S}chwer, {C}hair {PTC} 60/{V}\&{V} 10. {\it Eng Comput.} 2007\string;23(4)\string:245--252.
\newblock \url{https://doi.org/10.1007/s00366-007-0072-z}.

\bibitem{o2024implicit}
O’Hare TJ, Gourgiotis PA, Coombs WM, Augarde CE. An implicit {M}aterial {P}oint {M}ethod for micropolar solids undergoing large deformations. {\it Comput Methods Appl Mech Eng.} 2024\string;419\string:116668.
\newblock \url{https://doi.org/10.1016/j.cma.2023.116668}.

\end{thebibliography}

\appendix

\bmsection{Isothermal micropolar elastoplasticity derived from thermodynamics}\label{sec:appendixA}
The first law of thermodynamics in the isothermal case can be written as
\begin{equation}
    \begin{aligned}
    \dot{E}_{\text{total}} ~=~ P_{\text{input}}\,,
    \end{aligned}
    \label{eqn:first_law}
\end{equation}
where $E_{\text{total}}(t)$ is the total energy of the system and $P_{\text{input}}(t)$ is the mechanical power input due to external forces and moments. The total energy is defined as the kinetic and internal energy densities integrated over the body as
\begin{equation}
    \begin{aligned}
    E_{\text{total}} ~:= \int_{\Omega} \rho\left(k + \mathfrak{e}\right)\text{dv}\,,
    \end{aligned}
    \label{eqn:energy_total}
\end{equation}
where the kinetic energy per unit mass is defined as
\begin{equation}
    \begin{aligned}
    k ~:=~ \frac12\dot{u}_l\dot{u}_l \,+\, \frac12 j_{kl}\dot{\varphi}_k\dot{\varphi}_l\,.
    \end{aligned}
    \label{eqn:kinetic energy density}
\end{equation}
The mass density $\rho(\mathbf{x})$ and the microinertias $j_{kl}(\mathbf{x})$ are assumed to be time-invariant. The power input is defined using the body forces, body couples, and the surface tractions and moments integrated over the body
\begin{equation}
    \begin{aligned}
    P_{\text{input}} ~:= \int_{\Omega} \left(f_l\dot{u}_l + l_l\dot{\varphi}_l\right)\text{dv} \,+\, \int_{\partial\Omega} \left( t_{kl}n_k\dot{u}_l + m_{kl}n_k\dot{\varphi}_l \right) \text{da}\,.
    \end{aligned}
    \label{eqn:energy_total2}
\end{equation}
Now, due to the small deformation assumption, we can rewrite Eq\@.~\eqref{eqn:first_law} as
\begin{equation}
    \begin{aligned}
    \int_{\Omega} \left(\rho\ddot{u}_l\dot{u}_l \,+\, \rho j_{kl}\ddot{\varphi}_k\dot{\varphi}_l \,+\, \rho\dot{\mathfrak{e}} \,-\, f_l\dot{u}_l \,-\, l_l\dot{\varphi}_l\right)\text{dv} \,-\, \int_{\partial\Omega} \left( t_{kl}n_k\dot{u}_l + m_{kl}n_k\dot{\varphi}_l \right) \text{da} ~=~ 0\,.
    \end{aligned}
    \label{eqn:first_law_rewritten_1}
\end{equation}
Applying the quasi-static assumption (accelerations $\ddot{u}_k(\mathbf{x},t)$, $\ddot{\varphi}_k(\mathbf{x},t)$ are negligible, but the velocities $\dot{u}_k(\mathbf{x},t)$, $\dot{\varphi}_k(\mathbf{x},t)$ are not), Eq\@.~\eqref{eqn:first_law_rewritten_1} becomes
\begin{equation}
    \begin{aligned}
    \int_{\Omega} \left(\rho\dot{\mathfrak{e}} \,-\, f_l\dot{u}_l \,-\, l_l\dot{\varphi}_l\right)\text{dv} \,-\, \int_{\partial\Omega} \left( t_{kl}n_k\dot{u}_l + m_{kl}n_k\dot{\varphi}_l \right) \text{da} ~=~ 0\,.
    \end{aligned}
    \label{eqn:first_law_rewritten_2}
\end{equation}
By Divergence Theorem, Eq\@.~\eqref{eqn:first_law_rewritten_2} may be written with a single volume integral as
\begin{equation}
    \begin{aligned}
    \int_{\Omega} \left( \rho\dot{\mathfrak{e}} \,-\, f_l\dot{u}_l \,-\, l_l\dot{\varphi}_l - \left(t_{kl}\dot{u}_l\right)_{,k} \,-\, \left(m_{kl}\dot{\varphi}_l\right)_{,k} \right) \text{dv}  ~=~ 0\,.
    \end{aligned}
    \label{eqn:first_law_rewritten_3}
\end{equation}
Product rule for derivatives, and rearranging Eq\@.~\eqref{eqn:first_law_rewritten_3} gives
\begin{equation}
    \begin{aligned}
    \int_{\Omega} \left( \rho\dot{\mathfrak{e}} \,-\, t_{kl}\dot{u}_{l,k} \,-\, m_{kl}\dot{\varphi}_{l,k} \,-\, \left(t_{kl,k} + f_l\right)\dot{u}_l \,-\, \left(m_{kl,k} + l_l\right)\dot{\varphi}_l \right) \text{dv}  ~=~ 0\,.
    \end{aligned}
    \label{eqn:first_law_rewritten_4}
\end{equation}
Applying Eq\@.~\eqref{eqn:momemtum_balance} allows us to write Eq\@.~\eqref{eqn:first_law_rewritten_4} as 
\begin{equation}
    \begin{aligned}
    \int_{\Omega} \left( \rho\dot{\mathfrak{e}} \,-\, t_{kl}\left(\dot{u}_{l,k} \,+\, \epsilon_{lkm}\dot{\varphi}_{m} \right) \,-\, m_{kl}\dot{\varphi}_{l,k} \right) \text{dv}  ~=~ 0\,,
    \end{aligned}
    \label{eqn:first_law_rewritten_5}
\end{equation}
where then by applying Eq\@.~\eqref{eqn:strains} and localizing Eq\@.~\eqref{eqn:first_law_rewritten_5} from the arbitrariness of $\Omega$, we get the local form of the First Law as
\begin{equation}
    \begin{aligned}
    \rho\dot{\mathfrak{e}} ~=~ t_{kl}\dot{e}_{kl} \,+\, m_{kl}\dot{\rgamma}_{lk}\,.
    \end{aligned}
    \label{eqn:first_law_local_form}
\end{equation}

Next, we assume an elastoplastic additive decomposition of the strains
\begin{equation}
    \begin{aligned}
    e_{kl} ~=~ e_{kl}^{\text{e}} \,+\, e_{kl}^{\text{p}}\,,\quad\quad\quad \rgamma_{lk} ~=~ \rgamma_{lk}^{\text{e}} \,+\, \rgamma_{lk}^{\text{p}}\,,
    \end{aligned}
    \label{eqn:additive_decomposition}
\end{equation}
as well as the existence of a strain energy density function
\begin{equation}
    \begin{aligned}
    w\left(\mathbf{e}^{\text{e}},\boldsymbol{\rgamma}^{\text{e}},p,q\right) ~:=~ \rho\psi\left(\mathbf{e}^{\text{e}},\boldsymbol{\rgamma}^{\text{e}},p,q\right) ~:=~ \rho\mathfrak{e} \,-\, \rho\theta_0\eta\,,
    \end{aligned}
    \label{eqn:Helmholtz}
\end{equation}
where $\psi\left(\mathbf{e}^{\text{e}}(\mathbf{x},t),\boldsymbol{\rgamma}^{\text{e}}(\mathbf{x},t),p(\mathbf{x},t),q(\mathbf{x},t)\right)$ is the Helmholtz free energy per unit mass, $p(\mathbf{x},t)$ and $q(\mathbf{x},t)$ are equivalent cumulative plastic strains associated with hardening on the macro-level and micro-level, $\theta_0$ is the constant absolute temperature of the body, and $\eta(\mathbf{x},t)$ is the internal entropy per unit mass. Taking a time derivative of Eq\@.~\eqref{eqn:Helmholtz} we get 
\begin{equation}
    \begin{aligned}
    \dot{w}\left(\mathbf{e}^{\text{e}},\boldsymbol{\rgamma}^{\text{e}},p,q\right) ~=~ \frac{\partial w}{\partial e_{kl}^{\text{e}}}\dot{e}_{kl}^{\text{e}} \,+\, \frac{\partial w}{\partial \rgamma_{lk}^{\text{e}}}\dot{\rgamma}_{lk}^{\text{e}} \,+\, \frac{\partial w}{\partial p}\dot{p} \,+\, \frac{\partial w}{\partial q}\dot{q} ~=~ \rho\dot{\mathfrak{e}} \,-\, \rho\theta_0\dot{\eta}\,.
    \end{aligned}
    \label{eqn:Helmholtz_dot}
\end{equation}
Using Eq\@.~\eqref{eqn:first_law_local_form} and Eq\@.~\eqref{eqn:additive_decomposition}, we can write Eq\@.~\eqref{eqn:Helmholtz_dot} as 
\begin{equation}
    \begin{aligned}
    \rho\theta_0\dot{\eta} ~=~ \left(t_{kl} \,-\, \frac{\partial w}{\partial e_{kl}^{\text{e}}}\right)\dot{e}_{kl}^{\text{e}} \,+\, \left(m_{kl} \,-\, \frac{\partial w}{\partial \rgamma_{lk}^{\text{e}}}\right)\dot{\rgamma}_{lk}^{\text{e}} \,+\, t_{kl}\dot{e}_{kl}^{\text{p}} \,+\, m_{kl}\dot{\rgamma}_{lk}^{\text{p}} \,-\, \frac{\partial w}{\partial p}\dot{p} \,-\, \frac{\partial w}{\partial q}\dot{q}\,.
    \end{aligned}
    \label{eqn:theta_0_eta}
\end{equation}
The Second Law of Thermodynamics in the isothermal case can be written as 
\begin{equation}
    \begin{aligned}
    \dot{H} ~\geq~ 0 \,,
    \end{aligned}
    \label{eqn:second_law}
\end{equation}
where $H(t)$ is the internal entropy, defined as
\begin{equation}
    \begin{aligned}
    H ~:= \int_{\Omega} \rho\eta \text{dv} \,.
    \end{aligned}
    \label{eqn:internal_entropy}
\end{equation}
Due to the assumption of small deformations, and due to the arbitrariness of body $\Omega$, we can rewrite Eq\@.~\eqref{eqn:second_law} into the local form of the Second Law as
\begin{equation}
    \begin{aligned}
    \rho\dot{\eta} ~\geq~ 0 \,.
    \end{aligned}
    \label{eqn:second_law_local_form}
\end{equation}
Now, since $\theta_0$ is the absolute temperature of the body, we may use Eq\@.~\eqref{eqn:second_law_local_form} with Eq\@.~\eqref{eqn:theta_0_eta} to write the Clausius-Duhem, or dissipation inequality 
\begin{equation}
    \begin{aligned}
    \left(t_{kl} \,-\, \frac{\partial w}{\partial e_{kl}^{\text{e}}}\right)\dot{e}_{kl}^{\text{e}} \,+\, \left(m_{kl} \,-\, \frac{\partial w}{\partial \rgamma_{lk}^{\text{e}}}\right)\dot{\rgamma}_{lk}^{\text{e}} \,+\, t_{kl}\dot{e}_{kl}^{\text{p}} \,+\, m_{kl}\dot{\rgamma}_{lk}^{\text{p}} \,-\, \frac{\partial w}{\partial p}\dot{p} \,-\, \frac{\partial w}{\partial q}\dot{q} ~\geq~ 0\,.
    \end{aligned}
    \label{eqn:clausius_duhem}
\end{equation}
By the Coleman-Noll procedure, since $\dot{e}_{kl}^{\text{e}}(\mathbf{x},t)$ and $\dot{\rgamma}_{lk}^{\text{e}}(\mathbf{x},t)$ may be chosen independently, for Eq\@.~\eqref{eqn:clausius_duhem} to hold for all possible choices of these fields we must have 
\begin{equation}
    \begin{aligned}
    t_{kl} ~=~ \frac{\partial w}{\partial e_{kl}^{\text{e}}}\,,\quad\quad m_{kl} ~=~ \frac{\partial w}{\partial \rgamma_{lk}^{\text{e}}}\,,
    \end{aligned}
    \label{eqn:stresses_Helmholtz}
\end{equation}
giving us Eq\@.~\eqref{eqn:clausius_duhem} in the reduced dissipation inequality form as
\begin{equation}
    \begin{aligned}
     d\left(\mathbf{t},\mathbf{m}\right) \,:=~ t_{kl}\dot{e}_{kl}^{\text{p}} \,+\, m_{kl}\dot{\rgamma}_{lk}^{\text{p}} \,-\, \frac{\partial w}{\partial p}\dot{p} \,-\, \frac{\partial w}{\partial q}\dot{q} ~\geq~ 0\,,
    \end{aligned}
    \label{eqn:reduced_dissipation_ineq}
\end{equation}
where $d\left(\mathbf{t}(\mathbf{x},t),\mathbf{m}(\mathbf{x},t)\right)$ is called the mechanical dissipation at $(\mathbf{x},t)$ with units of power density. Integrating Eq\@.~\eqref{eqn:reduced_dissipation_ineq} over the body gives us the total mechanical dissipation at time $t$
\begin{equation}
    \begin{aligned}
     \mathcal{D}\left[\mathbf{t},\mathbf{m}\right](t) \,:=~ \int_{\Omega}d\left(\mathbf{t},\mathbf{m}\right)\text{dv} ~\geq~ 0\,,
    \end{aligned}
    \label{eqn:total_mech_diss_ineq}
\end{equation}

We now find the plastic flow rules via the Principle of Maximum Dissipation after defining multi-criteria yield conditions similar to \cite{forest2003elastoviscoplastic} as 
\begin{equation}
    \begin{aligned}
    f(\mathbf{t},p) ~\leq~ 0 \,,\quad\quad g(\mathbf{m},q) ~\leq~ 0\,,
    \end{aligned}
    \label{eqn:yield_conditions}
\end{equation}
where $f\left(\mathbf{t}(\mathbf{x},t),p(\mathbf{x},t)\right)$ and $g\left(\mathbf{m}(\mathbf{x},t),q(\mathbf{x},t)\right)$ are the yield functions of the macro-level and micro-level, respectively. Principle of Maximum Dissipation says that the ``true'' or physical stresses $t_{kl}(\mathbf{x},t)$ and couple stresses $m_{kl}(\mathbf{x},t)$ are those that maximize Eq\@.~\eqref{eqn:total_mech_diss_ineq} subjected to the inequality constraints imposed by Eq\@.~\eqref{eqn:yield_conditions}. Explicitly,
\begin{maxi}|l|
    {\textbf{t},\textbf{m}}{\mathcal{D}\left[\mathbf{t},\mathbf{m}\right](t)}{}{}
    \addConstraint{f(\mathbf{t},p) ~\leq~ 0}
    \addConstraint{g(\mathbf{m},q) ~\leq~ 0\,.}
    \label{eqn:max_dissipation}
\end{maxi}
This optimization problem can be reformulated with the Method of Lagrange Multipliers under inequality constraints by defining a functional
\begin{equation}
    \begin{aligned}
    D\left[\mathbf{t},\mathbf{m},\lambda_1,\lambda_2\right](t) \,:=~ \int_{\Omega}\bigg( d\left(\mathbf{t},\mathbf{m}\right) \,-\, \lambda_1f(\mathbf{t},p) \,-\, \lambda_2g(\mathbf{m},q) \bigg)\text{dv}\,,
    \end{aligned}
    \label{eqn:lagrangian}
\end{equation}
where $\lambda_1(\mathbf{x},t)\,,\lambda_2(\mathbf{x},t) \geq 0$ are the Lagrange multipliers, also called the plastic multipliers. The stationarity conditions for Eq\@.~\eqref{eqn:lagrangian} are given by
\begin{equation}
    \begin{aligned}
     \frac{\partial d}{\partial t_{kl}} \,-\, \lambda_1\frac{\partial f}{\partial t_{kl}} ~=~ 0 \,&,\quad\quad \frac{\partial d}{\partial m_{kl}} \,-\, \lambda_2\frac{\partial g}{\partial m_{kl}} ~=~ 0 \,,\\*[1mm]
    \lambda_1 ~\geq~ 0 \,&,\quad\quad \lambda_2 ~\geq~ 0 \,,\\*[1mm]
    f(\mathbf{t},p) ~\leq~ 0 \,&,\quad\quad g(\mathbf{m},q) ~\leq~ 0 \,,\\*[1mm]
    \lambda_1f(\mathbf{t},p) ~=~ 0 \,&,\quad\quad \lambda_2g(\mathbf{m},q) ~=~ 0 \,.
    \end{aligned}
    \label{eqn:KT_conditions}
\end{equation}
The first line of Eq\@.~\eqref{eqn:KT_conditions} gives the plastic flow rules as
\begin{equation}
    \begin{aligned}
    \dot{e}_{kl}^{\text{p}} ~=~ \lambda_1\frac{\partial f}{\partial t_{kl}} \,,\quad\,\,\,\,&\quad\quad\,\,\, \dot{\rgamma}_{lk}^{\text{p}} ~=~ \lambda_2\frac{\partial g}{\partial m_{kl}}\,,
    \end{aligned}
    \label{eqn:plastic_flow}
\end{equation}
while the last three lines of Eq\@.~\eqref{eqn:KT_conditions} give the loading/unloading conditions, also called the optimality, or Kuhn-Tucker conditions. Lastly, we give the consistency conditions as
\begin{equation}
    \begin{aligned}
    \lambda_1\dot{f}(\mathbf{t},p) ~=~ 0 \,,\quad\quad \lambda_2\dot{g}(\mathbf{m},q) ~=~ 0\,.
    \end{aligned}
    \label{eqn:consistency_conditions}
\end{equation}

\bmsection{Derivation of micropolar plastic strains at next time step}\label{sec:closedform}
Starting from Eq\@.~\eqref{eqn:stress_rates_flow_rules_plugged_in_EB_3_trial},
{
\begin{equation}
    \begin{aligned}
    t_{kl}  &~=~ t_{kl}^{\text{trial}} \,-\, \frac{p - p^{(n)}}{t_{\text{eq}}\left(\mathbf{t}\right)}\left(a_1A_{klmn}s_{(mn)} \,+\, a_2A_{klmn}s_{[mn]}\right)\,,\\*[1mm]
    m_{kl} &~=~ m_{kl}^{\text{trial}} \,-\, \frac{q - q^{(n)}}{m_{\text{eq}}(\mathbf{m})}\left(b_1B_{lkmn}m_{(nm)} \,+\, b_2B_{lkmn}m_{[nm]}\right)\,,
    \end{aligned}
    \label{eqn:stress_rates_flow_rules_plugged_in_EB_3_trial_appx}
\end{equation}
}
If we assume isotropy of the stiffness tensors in Eq\@.~\eqref{eqn:stress_rates_flow_rules_plugged_in_EB_3_trial_appx}, we may write
{
\begin{equation}
    \begin{aligned}
    t_{kl}  &~=~ t_{kl}^{\text{trial}} \,-\, \frac{p - p^{(n)}}{t_{\text{eq}}\left(\mathbf{t}\right)}\left(2a_1\mu s_{(kl)} \,+\, 2a_2\kappa s_{[kl]}\right)\,,\\*[1mm]
    m_{kl} &~=~ m_{kl}^{\text{trial}} \,-\, \frac{q - q^{(n)}}{m_{\text{eq}}(\mathbf{m})}\left(b_1\alpha m_{nn}\delta_{kl} \,+\, b_1(\beta + \gamma)m_{(lk)} \,+\, b_2(\beta - \gamma)m_{[lk]}\right)\,.
    \end{aligned}
    \label{eqn:stress_rates_flow_rules_plugged_in_isotropic}
\end{equation}
}
Let us assume for simplicity $\alpha(\mathbf{x}) = 0$ $\forall\mathbf{x}\in\Omega$, which is energetically admissible by Eq\@.~\eqref{eqn:energetic_bounds}.
Since $s_{kk} = s_{(kk)} = s_{[kk]} = 0$, we have $t_{kk} = t_{kk}^{\text{trial}}$, so by subtracting $\frac13 t_{nn}\delta_{kl}$ from both sides of the first line of Eq\@.~\eqref{eqn:stress_rates_flow_rules_plugged_in_isotropic}, we can write 
{
\begin{equation}
    \begin{aligned}
    s_{kl}  &~=~ s_{kl}^{\text{trial}} \,-\, \frac{p - p^{(n)}}{t_{\text{eq}}\left(\mathbf{t}\right)}\left(2a_1\mu s_{(kl)} \,+\, 2a_2\kappa s_{[kl]}\right)\,,\\*[1mm]
    m_{kl} &~=~ m_{kl}^{\text{trial}} \,-\, \frac{q - q^{(n)}}{m_{\text{eq}}(\mathbf{m})}\left(b_1(\gamma + \beta)m_{(kl)} \,+\, b_2(\gamma - \beta)m_{[kl]}\right)\,.
    \end{aligned}
    \label{eqn:stress_rates_flow_rules_plugged_in_isotropic_dev}
\end{equation}
}
One may decompose Eq\@.~\eqref{eqn:stress_rates_flow_rules_plugged_in_isotropic_dev} into symmetric and skew-symmetric parts
{
\begin{equation}
    \begin{aligned}
    s_{(kl)}  &~=~ s_{(kl)}^{\text{trial}} \,-\, \frac{2\left(p - p^{(n)}\right)a_1\mu}{t_{\text{eq}}\left(\mathbf{t}\right)} s_{(kl)}\,,\\*[1mm]
    s_{[kl]}  &~=~ s_{[kl]}^{\text{trial}} \,-\, \frac{2\left(p - p^{(n)}\right)a_2\kappa}{t_{\text{eq}}\left(\mathbf{t}\right)} s_{[kl]}\,,\\*[1mm]
    m_{(kl)}  &~=~ m_{(kl)}^{\text{trial}} \,-\, \frac{\left(q - q^{(n)}\right)b_1(\gamma + \beta)}{m_{\text{eq}}\left(\mathbf{m}\right)} m_{(kl)}\,,\\*[1mm]
    m_{[kl]}  &~=~ m_{[kl]}^{\text{trial}} \,-\, \frac{\left(q - q^{(n)}\right)b_2(\gamma - \beta)}{m_{\text{eq}}\left(\mathbf{m}\right)} m_{[kl]}\,.
    \end{aligned}
    \label{eqn:stress_rates_flow_rules_plugged_in_isotropic_dev_decomp}
\end{equation}
}
This form is advantageous because the unknown stress quantities are directly proportional to the known trial quantities. 
Explicitly, 
{
\begin{equation}
    \begin{aligned}
    s_{(kl)}^{\text{trial}} &~=~ \left(\frac{t_{\text{eq}}(\mathbf{t}) \,+\, 2\left(p - p^{(n)}\right)a_1\mu}{t_{\text{eq}}(\mathbf{t})}\right)s_{(kl)} \,,\\*[1mm]
    s_{[kl]}^{\text{trial}} &~=~ \left(\frac{t_{\text{eq}}(\mathbf{t}) \,+\, 2\left(p - p^{(n)}\right)a_2\kappa}{t_{\text{eq}}(\mathbf{t})}\right)s_{[kl]} \,,\\*[1mm]
    m_{(kl)}^{\text{trial}} &~=~ \left(\frac{m_{\text{eq}}(\mathbf{m}) \,+\, \left(q - q^{(n)}\right)b_1(\gamma + \beta)}{m_{\text{eq}}(\mathbf{m})}\right)m_{(kl)}\,,\\*[1mm]
    m_{[kl]}^{\text{trial}} &~=~ \left(\frac{m_{\text{eq}}(\mathbf{m}) \,+\, \left(q - q^{(n)}\right)b_2(\gamma - \beta)}{m_{\text{eq}}(\mathbf{m})}\right)m_{[kl]}\,.
    \end{aligned}
\label{eqn:stress_rates_flow_rules_plugged_in_isotropic_dev_decompp}
\end{equation}
}
Therefore, when $p(\mathbf{x})$ and $q(\mathbf{x})$ at time $t_{n+1}$ are known, we may solve for the unknown stresses $t_{kl}(\mathbf{x})$ and $m_{kl}(\mathbf{x})$ at time $t_{n+1}$.
These equations will be used when plastic flow is happening, so $t_{\text{eq}}(\mathbf{t}(\mathbf{x}))$ and $m_{\text{eq}}(\mathbf{m}(\mathbf{x}))$ at time $t_{n+1}$ are determined by the yield functions, Eqs\@.~\eqref{eqn:yield_function_f} and~\eqref{eqn:yield_function_g}.
To solve for $p(\mathbf{x})$ and $q(\mathbf{x})$ at time $t_{n+1}$ in closed form, we compute $t_{\text{eq}}(\mathbf{t}^{\text{trial}}(\mathbf{x}))$ and $m_{\text{eq}}(\mathbf{m}^{\text{trial}}(\mathbf{x}))$ using Eq\@.~\eqref{eqn:stress_rates_flow_rules_plugged_in_isotropic_dev_decompp}. $t_{\text{eq}}(\mathbf{t}^{\text{trial}}(\mathbf{x}))$ first,
{
\begin{equation}
    \begin{aligned}
    t_{\text{eq}}(\mathbf{t}^{\text{trial}}) &~=~ \sqrt{a_1s_{(kl)}^{\text{trial}}s_{(kl)}^{\text{trial}} \,+\, a_2s_{[kl]}^{\text{trial}}s_{[kl]}^{\text{trial}}}\,,\\*[1mm]
    &~=~ t_{\text{eq}}(\mathbf{t}) + 2\left(p - p^{(n)}\right)a_1\mu\,.
    \end{aligned}
    \label{eqn:equivalent_trial_stress}
\end{equation}
}
where we have assumed that
\begin{equation}
    \begin{aligned}
    a_2 = a_1\frac{\mu}{\kappa}\,,
    \end{aligned}
    \label{eqn:equivalent_trial_stress_assumption}
\end{equation}
for all $\mathbf{x}\in\Omega$.
For $m_{\text{eq}}(\mathbf{m}^{\text{trial}}(\mathbf{x}))$, we have
{
\begin{equation}
    \begin{aligned}
    m_{\text{eq}}(\mathbf{m}^{\text{trial}}) &~=~ \sqrt{b_1m_{(kl)}^{\text{trial}}m_{(kl)}^{\text{trial}} \,+\, b_2m_{[kl]}^{\text{trial}}m_{[kl]}^{\text{trial}}}\,,\\*[1mm]
    &~=~ m_{\text{eq}}(\mathbf{m}) + \left(q - q^{(n)}\right)b_1(\gamma + \beta)\,,
    \end{aligned}
    \label{eqn:equivalent_trial_couple_stress}
\end{equation}
}
where an assumption similar to Eq\@.~\eqref{eqn:equivalent_trial_stress_assumption} has been made, such that:
\begin{equation}
    \begin{aligned}
    b_2 = b_1\frac{\gamma + \beta}{\gamma - \beta}\,.
    \end{aligned}
    \label{eqn:equivalent_trial_couple_stress_assumption}
\end{equation}
for all $\mathbf{x}\in\Omega$.
We can now solve for $p(\mathbf{x})$ and $q(\mathbf{x})$ at time $t_{n+1}$ using Eqs\@.~\eqref{eqn:equivalent_trial_stress} and~\eqref{eqn:equivalent_trial_couple_stress} along with $f(\mathbf{t}(\mathbf{x}),p(\mathbf{x})) = 0$, $g(\mathbf{m}(\mathbf{x}),q(\mathbf{x})) = 0$, and Eqs\@.~\eqref{eqn:yield_function_f}-\eqref{eqn:radius_2} to get
{
\begin{equation}
    \begin{aligned}
    p &~=~ p^{(n)}\frac{2a_1\mu}{t_{\text{H}} + 2a_1\mu} \,+\, \frac{t_{\text{eq}}(\mathbf{t}^{\text{trial}}) - t_{\text{Y}}}{t_{\text{H}} + 2a_1\mu}\,,\\*[1mm]
    q &~=~ q^{(n)}\frac{b_1(\gamma + \beta)}{m_{\text{H}} + b_1(\gamma + \beta)} \,+\, \frac{m_{\text{eq}}(\mathbf{m}^{\text{trial}}) - m_{\text{Y}}}{m_{\text{H}} + b_1(\gamma+\beta)}\,.
    \end{aligned}
    \label{eqn:trial_stresses2_appx}
\end{equation}
}

\bmsection{Micropolar elastoplastic continuum tangents}\label{sec:continuum_tangents}
While not used explicitly in the algorithms presented here, it is useful for verification purposes to have the continuum tangents. During plastic loading, we have $\lambda_1,\lambda_2>0$, so Eq\@.~\eqref{eqn:consistency_conditions} can be written after applying chain rule as
\begin{equation}
    \begin{aligned}
    \dot{f}(\mathbf{t},p) ~=~ \frac{\partial f}{\partial t_{kl}}\dot{t}_{kl} \,+\, \frac{\partial f}{\partial p}\dot{p} ~=~ 0 \,,\quad\quad \dot{g}(\mathbf{m},q) ~=~ \frac{\partial g}{\partial m_{kl}}\dot{m}_{kl} \,+\, \frac{\partial g}{\partial q}\dot{q} ~=~ 0\,.
    \end{aligned}
    \label{eqn:consistency_conditions_}
\end{equation}
Plugging in stress rates and plastic flow rules, Eqs\@.~\eqref{eqn:stress_rates},~\eqref{eqn:plastic_flow},~\eqref{eqn:p_dot},~\eqref{eqn:q_dot}, and rearranging, we get the following two equations for $\dot{p}(\mathbf{x},t)$ and $\dot{q}(\mathbf{x},t)$
\begin{equation}
    \begin{aligned}
    \dot{p} ~=~ \frac{\frac{\partial f}{\partial t_{kl}}A_{klmn}\dot{e}_{mn}}{\frac{\partial f}{\partial t_{ab}}A_{abcd}\frac{\partial f}{\partial t_{cd}} - \frac{\partial f}{\partial p}}\,,\quad\quad \dot{q} ~=~ \frac{\frac{\partial g}{\partial m_{kl}}B_{lkmn}\dot{\rgamma}_{mn}}{\frac{\partial g}{\partial m_{ba}}B_{abcd}\frac{\partial g}{\partial m_{dc}} - \frac{\partial g}{\partial q}}\,.
    \end{aligned}
    \label{eqn:p_dot_and_q_dot}
\end{equation}
Using Eq\@.~\eqref{eqn:p_dot_and_q_dot} in the stress rate equations, we arrive at 
\begin{equation}
    \begin{aligned}
    \dot{t}_{kl} ~=~ \left(A_{klmn} - \frac{A_{klrs}\frac{\partial f}{\partial t_{rs}}\frac{\partial f}{\partial t_{pq}}A_{pqmn}}{\frac{\partial f}{\partial t_{ab}}A_{abcd}\frac{\partial f}{\partial t_{cd}} - \frac{\partial f}{\partial p}}\right)\dot{e}_{mn}\,,\quad\quad \dot{m}_{kl} ~=~ \left(B_{lkmn} - \frac{B_{lkrs}\frac{\partial g}{\partial m_{sr}}\frac{\partial g}{\partial m_{pq}}B_{qpmn}}{\frac{\partial g}{\partial m_{ba}}B_{abcd}\frac{\partial g}{\partial m_{dc}} - \frac{\partial g}{\partial q}}\right)\dot{\rgamma}_{mn}\,,
    \end{aligned}
    \label{eqn:plastic_stress_rates}
\end{equation}
so now we may write the micropolar elastoplastic continuum tangent moduli as 
\begin{equation}
    \begin{aligned}
    A_{klmn}^{\text{ep}}(\mathbf{t},p) ~=~ \begin{cases}
    A_{klmn} & f(\mathbf{t},p) < 0 \\
    A_{klmn} - \frac{A_{klrs}\frac{\partial f}{\partial t_{rs}}\frac{\partial f}{\partial t_{pq}}A_{pqmn}}{\frac{\partial f}{\partial t_{ab}}A_{abcd}\frac{\partial f}{\partial t_{cd}} - \frac{\partial f}{\partial p}} & f(\mathbf{t},p) = 0\,,
    \end{cases}
    \end{aligned}
    \label{eqn:continuum_tangent_A}
\end{equation}
and
\begin{equation}
    \begin{aligned}
    B_{lkmn}^{\text{ep}}(\mathbf{m},q) ~=~ \begin{cases}
    B_{lkmn} & g(\mathbf{m},q) < 0 \\
    B_{lkmn} - \frac{B_{lkrs}\frac{\partial g}{\partial m_{sr}}\frac{\partial g}{\partial m_{pq}}B_{qpmn}}{\frac{\partial g}{\partial m_{ba}}B_{abcd}\frac{\partial g}{\partial m_{dc}} - \frac{\partial g}{\partial q}} & g(\mathbf{m},q) = 0\,.
    \end{cases}
    \end{aligned}
    \label{eqn:continuum_tangent_B}
\end{equation}
The stress rate equations can then be written as 
\begin{equation}
    \begin{aligned}
    \dot{t}_{kl} ~=~ A_{klmn}^{\text{ep}}\dot{e}_{mn} \,,\quad\quad \dot{m}_{kl} ~=~ B_{lkmn}^{\text{ep}}\dot{\rgamma}_{mn} \,,
    \end{aligned}
    \label{eqn:stress_rates_with_continuum_tangent}
\end{equation}
and, following \cite{simo2006computational}, Eq\@.~\eqref{eqn:p_dot_and_q_dot} for elastoplasticity is given by 
\begin{equation}
    \begin{aligned}
    \dot{p} ~=~ \frac{\max\left(\frac{\partial f}{\partial t_{kl}}A_{klmn}\dot{e}_{mn}\,,0\right)}{\frac{\partial f}{\partial t_{ab}}A_{abcd}\frac{\partial f}{\partial t_{cd}} - \frac{\partial f}{\partial p}}\,,\quad\quad \dot{q} ~=~ \frac{\max\left(\frac{\partial g}{\partial m_{kl}}B_{lkmn}\dot{\rgamma}_{mn}\,,0\right)}{\frac{\partial g}{\partial m_{ba}}B_{abcd}\frac{\partial g}{\partial m_{dc}} - \frac{\partial g}{\partial q}}\,.
    \end{aligned}
    \label{eqn:p_dot_and_q_dot_EP}
\end{equation}

\bmsection{Verification}\label{sec:verification}
{
Following the procedure from Schwer \cite{schwer2007overview}, we verified our implementation of Alg\@.~\ref{alg:radial_return} and Alg\@.~\ref{alg:fft} in two steps:
1) code verification (Appendix~\ref{sec:codeverif}), and
2) calculation verification (Appendix~\ref{sec:codecalc}).
\textit{Code verification} aims at establishing confidence that the model (micropolar elastoplasticity) and solution algorithms (Alg\@.~\ref{alg:radial_return} and Alg\@.~\ref{alg:fft}) are correct, usually by comparing to an analytical solution when available, or alternatively using the method of manufactured solutions (MMS).
\textit{Calculation verification} on the other hand, aims at establishing confidence that the discrete numerical solution of the model is accurate, by means of a mesh-refinement study to estimate discretization errors.
} 
\bmsubsection{Code verification}\label{sec:codeverif}
{
In this subsection, we verified the implemented 3D micropolar elastoplasticity code using an extension of MMS, called the method of numerically manufactured solutions (MNMS), introduced by Chen et al. \cite{chen2017proposal}.
This extension is necessary in the case of elastoplasticity because of the discontinuous nature of the constitutive equations, see Eqs\@.~\eqref{eqn:continuum_tangent_A} and~\eqref{eqn:continuum_tangent_B}.
For micropolar elastoplasticity, the basic idea of MNMS (specifically MNMS-II in \cite{chen2017proposal}) is to
(a) manufacture (total) strains $e_{kl}^{\text{m}}(\mathbf{x},t)$, $\rgamma_{kl}^{\text{m}}(\mathbf{x},t)$,
(b) numerically integrate in time Eqs\@.~\eqref{eqn:stress_rates_with_continuum_tangent} and~\eqref{eqn:p_dot_and_q_dot_EP} at fixed values of $\mathbf{x}$ in order to construct the numerically manufactured polarization stress fields $\tau_{kl}^{\text{m}}(\mathbf{x},t)$, $\mu_{kl}^{\text{m}}(\mathbf{x},t)$, and
(c) run Alg\@.~\ref{alg:fft} with the first line inside the while-loop rewritten as
\begin{equation}
    \begin{aligned}
    \tau_{kl}^{(i)} ~=~ t_{kl}^{(n,i)} \,-\, A_{klmn}^0 e_{mn}^{(n,i)} \,-\, \tau_{kl}^{\text{m}(n)}\,,\quad\quad \mu_{kl}^{(i)} ~=~ m_{kl}^{(n,i)} \,-\, B_{lkmn}^0 \rgamma_{mn}^{(n,i)} \,-\, \mu_{kl}^{\text{m}(n)}\,.
    \end{aligned}
    \label{eqn:manufactured_polarization_stress}
\end{equation}
Upon completion of Alg\@.~\ref{alg:fft}, it should hold that $\big(e_{kl}(\mathbf{x},t)\,, \rgamma_{kl}(\mathbf{x},t)\,,t_{kl}(\mathbf{x},t)\,, m_{kl}(\mathbf{x},t)\big)$ is equal (within certain tolerance) to $\big(e_{kl}^{\text{m}}(\mathbf{x},t)\,, \rgamma_{kl}^{\text{m}}(\mathbf{x},t)\,,\tau_{kl}^{\text{m}}(\mathbf{x},t)\,, \mu_{kl}^{\text{m}}(\mathbf{x},t)\big)$.
}

{
Taking inspiration from the manufactured solution in \cite{o2024implicit}, we write ours here as 
\begin{equation}
    \begin{aligned}
    u_i^{\text{m}}(\mathbf{x},t) ~&=~ t\sin(2\pi x_1)\sin(2\pi x_2)\sin(2\pi x_3) \,+\, E_{ij}^{\text{m}}(t)x_j\,,\\*[1mm]
    \varphi_i^{\text{m}}(\mathbf{x},t) ~&=~ t\sin(2\pi x_1)\sin(2\pi x_2)\sin(2\pi x_3) \,+\, \Gamma_{ij}^{\text{m}}(t)x_j\,,
    \end{aligned}
    \label{eqn:manufactured_displ_solutions}
\end{equation}
where, in order to test each component, the manufactured average strains are arbitrarily set to
\begin{equation}
    \begin{aligned}
    \left[E_{ij}^{\text{m}}(t)\right] ~=~ t\begin{bmatrix}
    1.0 & 2.0 & 3.0 \\
    4.0 & 5.0 & 6.0 \\
    7.0 & 8.0 & 9.0 
    \end{bmatrix}\,,\quad\quad 
    \left[\Gamma_{ij}^{\text{m}}(t)\right] ~=~ t\begin{bmatrix}
    1.5 & 2.5 & 3.5 \\
    4.5 & 5.5 & 6.5 \\
    7.5 & 8.5 & 9.5 
    \end{bmatrix}\,.
    \end{aligned}
    \label{eqn:manufactured_avg_strains}
\end{equation}
Using Eq\@.~\eqref{eqn:strains} and taking a time derivative of Eq\@.~\eqref{eqn:manufactured_displ_solutions}, we get $\dot{e}_{kl}^{\text{m}}(\mathbf{x},t)$ and $\dot{\rgamma}_{kl}^{\text{m}}(\mathbf{x},t)$.
Before being able to numerically integrate to get $\tau_{kl}^{\text{m}}(\mathbf{x},t)$ and $\mu_{kl}^{\text{m}}(\mathbf{x},t)$, we must specify the constitutive parameters and geometry of the unit cell.
For the geometry, we adopted a $4\times 4\times 4$ voxelized cubic composite.
Material phase 1 was set as the inner centered $2\times 2\times 2$ cube and material phase 2 as the outer layer of voxels surrounding phase 1.
For the constitutive parameters, we specialized to the case of perfect plasticity, such that $t_{\text{H}}(\mathbf{x}) = 0$ and $m_{\text{H}}(\mathbf{x}) = 0$ for all $\mathbf{x}\in\Omega$, and $\alpha(\mathbf{x}) = 0$ for all $\mathbf{x}\in\Omega$ (see Eq\@.~\eqref{eqn:stress_rates_flow_rules_plugged_in_isotropic}).
The other constitutive parameters were chosen to satisfy Eq\@.~\eqref{eqn:energetic_bounds} and to represent a contrast between phases.
The properties of phase 1 properties were set to:
$\lambda^1 = \mu^1 = 1.0$,
$\kappa^1 = 0.5$,
$\alpha^1 = 0.0$,
$\beta^1 = 0.5$,
$\gamma^1 = 1.0$,
$t_{\text{Y}}^1 = 4.0$,
$t_{\text{H}}^1 = 0.0$,
$a_1^1 = 1.5$,
$m_{\text{Y}}^1 = 4.0$,
$m_{\text{H}}^1 = 0.0$, and
$b_1^1 = 1.5$;
and those of phase 2 were set to:
$\lambda^2 = \mu^2 = 1.5$,
$\kappa^2 = 0.75$,
$\alpha^2 = 0.0$,
$\beta^2 = 0.25$,
$\gamma^2 = 1.5$,
$t_{\text{Y}}^2 = 4.5$,
$t_{\text{H}}^2 = 0.0$,
$a_1^2 = 1.5$,
$m_{\text{Y}}^2 = 4.5$,
$m_{\text{H}}^2 = 0.0$, and
$b_1^2 = 1.5$.
The cubic side length was set to $L=1$.
Numerical time integration of Eqs\@.~\eqref{eqn:stress_rates_with_continuum_tangent} and~\eqref{eqn:p_dot_and_q_dot_EP} was performed with Mathematica using the ``NDSolve'' function with an ``AccuracyGoal'' set to 12 (corresponding roughly to the number of digits of accuracy sought in the final result).
We integrated from $t_1 = 0.0$ to $t_{101} = 1.0$, with the number of time steps $N = 100$ and $\Delta t = 0.01$.
The resultants $\tau_{kl}^{\text{m}}(\mathbf{x},t)$ and $\mu_{kl}^{\text{m}}(\mathbf{x},t)$ were then fed to Alg\@.~\ref{alg:fft} via Eq\@.~\eqref{eqn:manufactured_polarization_stress}, and by using FFT error measure Eq\@.~\eqref{eqn:err_local} with an $\varepsilon = 1\times10^{-9}$, we obtained the output tensor fields for stresses and strains  $t_{kl}(\mathbf{x},t)$, $m_{kl}(\mathbf{x},t)$, $e_{kl}(\mathbf{x},t)$, $\rgamma_{kl}(\mathbf{x},t)$.
The absolute errors per voxel over time between each component of the numerical output and the manufactured solutions are given in Fig\@.~\ref{fig:MNMS}.
Each color corresponds to a different component of a tensor (errors corresponding to $t_{11}$, $e_{11}$, $m_{11}$, and $\rgamma_{11}$ will all be red, for example), and within each color there are $4\times 4\times 4 = 64$ lines plotted that correspond to each voxel.
The main result is that the maximum error between the output of Alg\@.~\ref{alg:fft} and the manufactured solutions is below the error threshold of the FFT-based method, $\varepsilon = 1\times10^{-9}$.
Hence, these results demonstrate that the code verification of our implementation of Alg\@.~\ref{alg:radial_return} and Alg\@.~\ref{alg:fft} has been achieved.
}

\bmsubsection{Calculation verification}\label{sec:codecalc}
{
We show in Fig\@.~\ref{fig:convergence} a convergence study for two different geometries (a laminate with 50\% volume fraction, and a centered spherical inclusion with about 6.5\% volume fraction).
This plot shows the absolute error of the spatially averaged $t_{12}$ component of the stress tensor at the last time step, where the adopted reference is the most resolved solution in space and time, shown as the white ``reference'' square in the top right corner of the figure.
These geometries are chosen since the laminate is exactly represented for all resolutions considered, whereas the spherical inclusion is never perfectly represented by the voxels, due to it's curved surface.
The elastic material parameters are the same as the last section, but the plastic parameters are now
$t_{\text{Y}}^1 = 1.0$,
$t_{\text{H}}^1 = 0.0$,
$a_1^1 = 1.5$,
$m_{\text{Y}}^1 = 1.0$,
$m_{\text{H}}^1 = 0.0$, and
$b_1^1 = 1.5$ for phase 1 and
$t_{\text{Y}}^2 = 1.5$,
$t_{\text{H}}^2 = 0.0$,
$a_1^2 = 1.5$,
$m_{\text{Y}}^2 = 1.5$,
$m_{\text{H}}^2 = 0.0$, and
$b_1^2 = 1.5$ for phase 2.
The cubic side length is set as $L=1$.
Like before, we integrate from $t_1 = 0.0$ to $1.0$, and the only nonzero component of average strain rate imposed is $\dot{E}_{12}(t) = 1$.
The error metric used is Eq\@.~\eqref{eqn:err_average} with $\varepsilon = 1\times 10^{-4}$.
For the laminate, we see that increasing the spatial resolution has little effect on the solution until we refine the temporal resolution; i.e., the temporal resolution carries the majority of the difference.
For the spherical inclusion, we observe the opposite.
The spatial resolution matters the most because the more voxels were used, the closer the image approximates a spherical geometry.
This causes lower resolution solutions to be farther from reference than the corresponding solutions in the laminate.
}
\setcounter{figure}{0}
\begin{figure*}[ht]
    \centering
    \includegraphics[width=0.8\textwidth]{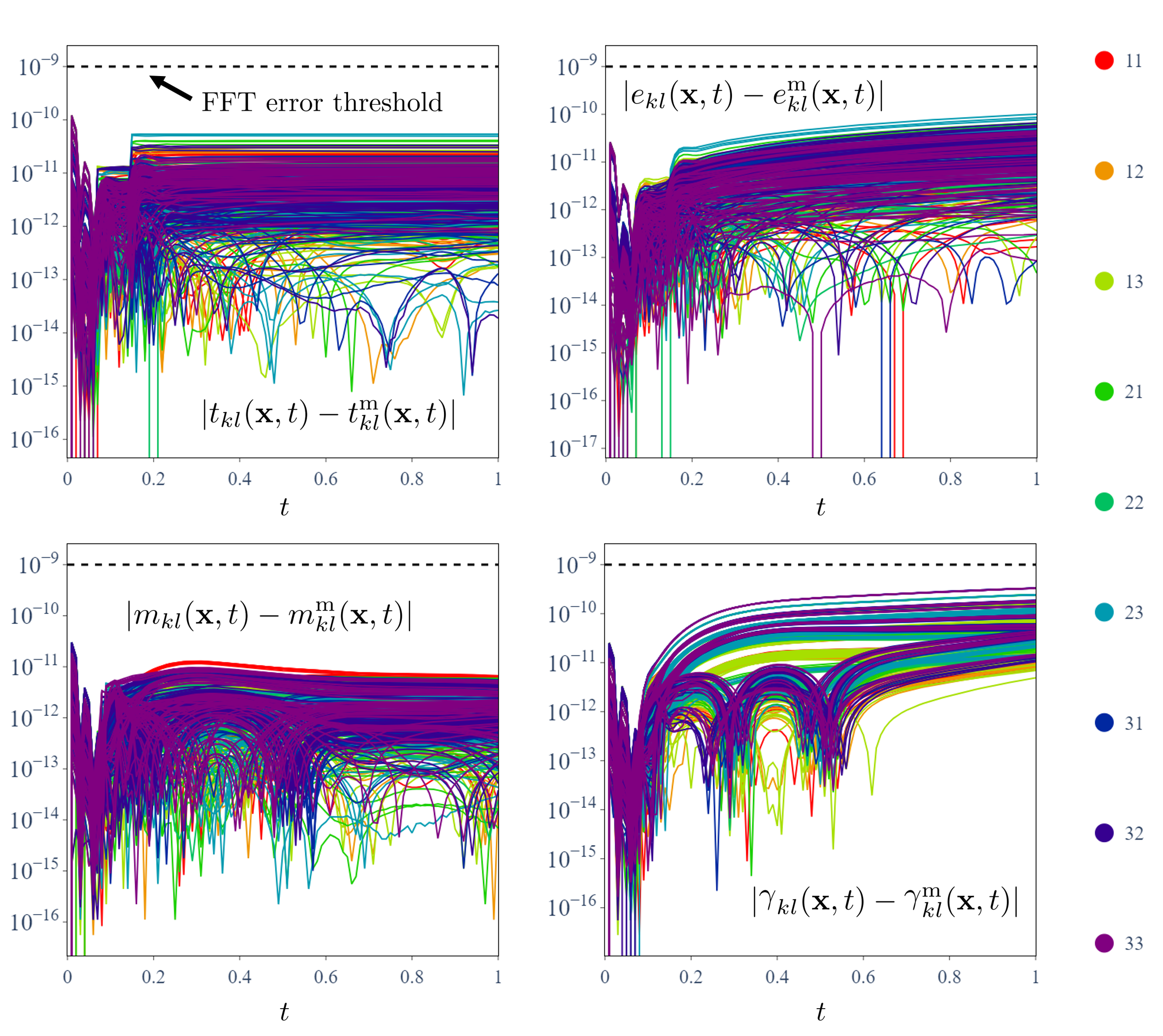}
    \caption[]{Absolute errors per voxel over time between each component of the FFT output and the manufactured solutions. Different colors correspond to different tensor components (see legend on right). Different lines of the same color correspond to the $4\times 4\times 4 = 64$ voxels that were simulated over time. All errors are bounded by the FFT error threshold of $\varepsilon = 1\times10^{-9}$.} 
    \label{fig:MNMS}
\end{figure*}

\begin{figure*}[ht]
    \centering
    \includegraphics[width=0.8\textwidth]{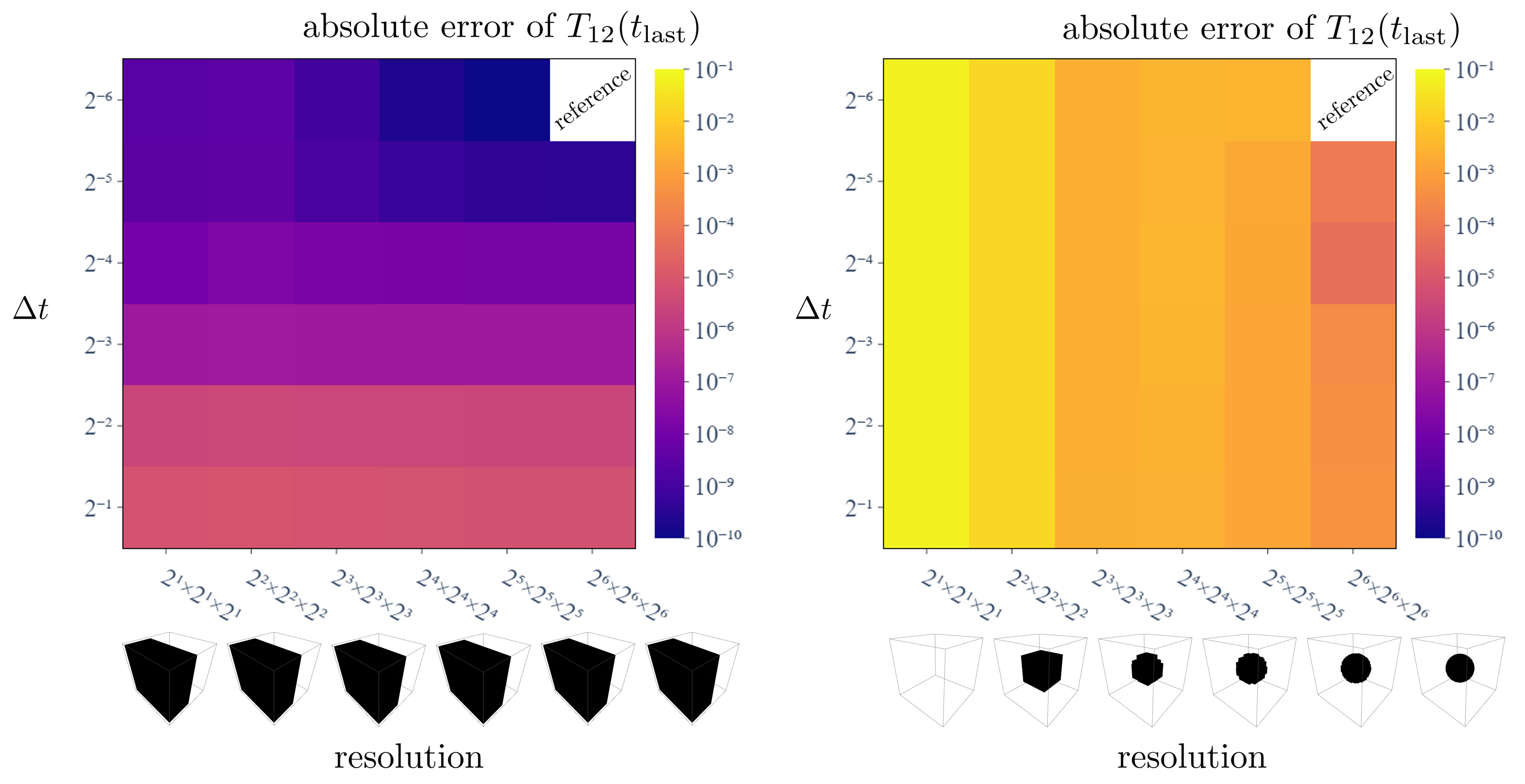}
    \caption[]{Convergence study for two a geometries: (left) laminate with 50\% volume fraction, (right) centered spherical inclusion with radius $1/4$ the cubic side length. Color represents the absolute error of the spatially averaged $t_{12}$ component of the stress tensor at the last time step, where reference is taken as the most resolved solution in space and time, shown as the white ``reference'' square in the top right. The laminate geometry is exactly represented in voxels for all resolutions considered; the spherical inclusion is never perfectly represented by voxels, due to the curvature of it's surface.} 
    \label{fig:convergence}
\end{figure*}

\end{document}